%% file: Thesis_final.tex
\documentclass[a4paper,12pt,onecolumn]{report}
\usepackage{latexsym}
\usepackage{algorithmic}
\usepackage{algorithm}
\usepackage{appendix}
\usepackage{amssymb}
\usepackage{epsfig}
\usepackage[dvips]{color}
\usepackage{graphicx}
\usepackage{amsmath}
\usepackage{url}
\usepackage{array}
\usepackage{subfigure}
\usepackage{setspace}
\usepackage{geometry}
\usepackage[colorlinks=true, pdfstartview=FitV, linkcolor=blue, citecolor=red, urlcolor=blue]{hyperref}
\makeatletter
\newcommand{\LyX}{L\kern-.1667em\lower.25em\hbox{Y}\kern-.125emX\spacefactor1000}
\makeatother

\begin{document}
\begin{titlepage}
\thispagestyle{empty}
\centering
\hfill \\
{\Large{\textbf{M. Tech. Degree Dissertation}}}
\hfill \\
\hfill \\
\hfill \\
{\Huge{\textbf{An Energy-Aware On-Demand}}}
\hfill \\
\hfill \\
{\Huge{\textbf{Routing Protocol}}}
\hfill \\
\hfill \\
{\Huge{\textbf{for Ad-Hoc Wireless Networks}}}
\hfill \\
\hfill \\
\hfill \\
{\large{Submitted in partial fulfillment of the requirements}}
\hfill \\
{\large{for the degree of}}
\hfill \\
\hfill \\
\hfill \\
{\large{\textbf{Master of Technology}}}
\hfill \\
{\textit{in}}
\hfill \\
{\large{\textit{Communication Engineering and Signal Processing}}}
\hfill \\
\hfill \\
\hfill \\
{\textit{by}}
\hfill \\
{\large \textbf{Mallapur Veerayya}}
\hfill \\
{\large \textbf{Roll No. 06307416}}
\hfill \\
\hfill \\
\hfill \\
{\itshape Under the guidance of}
\hfill \\
\hfill \\
{\bfseries \Large Prof. Abhay Karandikar}
\hfill \\
\hfill \\
{\bfseries \Large Dr. Vishal Sharma, Metanoia, Inc.}
\hfill \\
\hfill \\
\begin{figure}[ht]
\centering
\includegraphics[scale = 0.30]{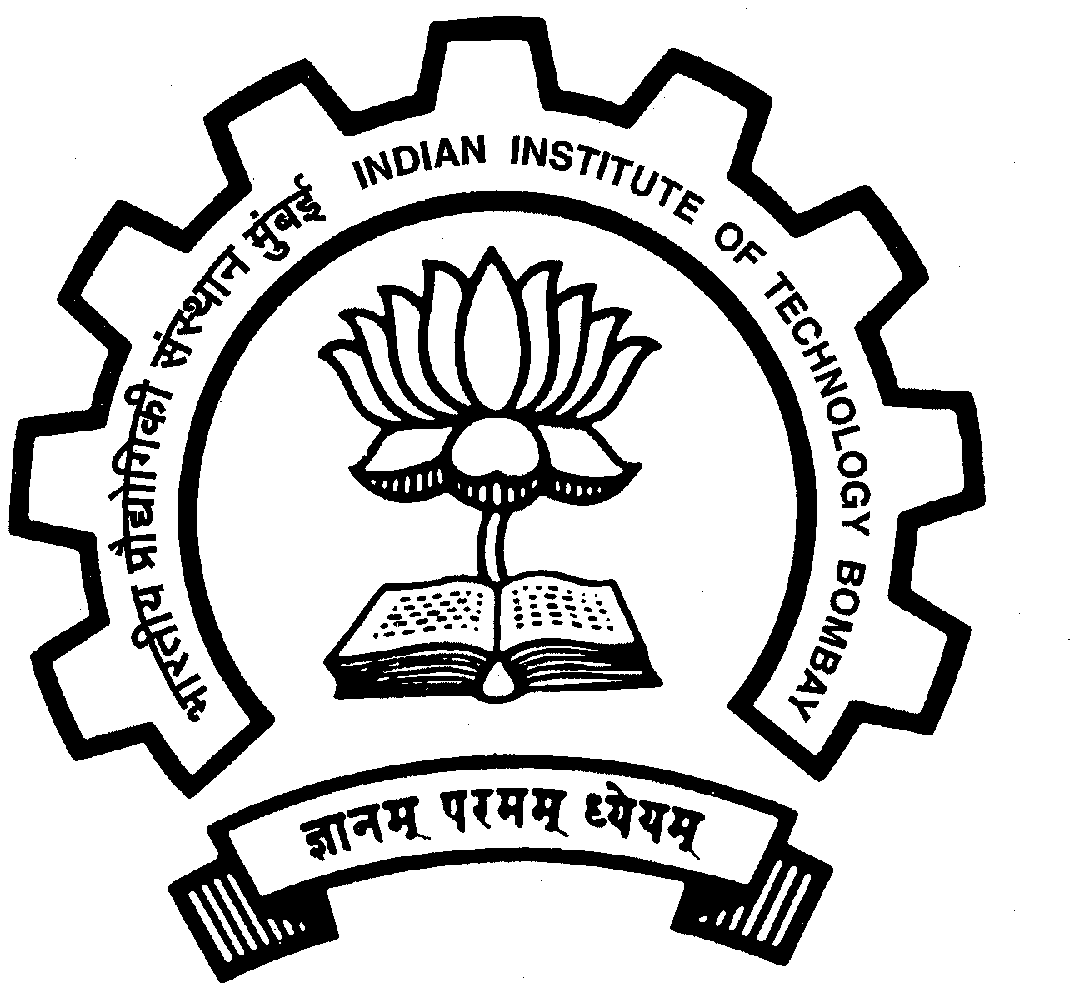}
\end{figure}
Department of Electrical Engineering\\
Indian Institute of Technology Bombay\\
July, 2008
\end{titlepage}
\include{ack}
\include{abs}
\pagebreak
\pagenumbering{roman}
\tableofcontents
\cleardoublepage
\phantomsection
\addcontentsline{toc}{chapter}{List of Tables}
\listoftables
\cleardoublepage
\phantomsection
\addcontentsline{toc}{chapter}{List of Figures}
\listoffigures
\pagebreak
\pagenumbering{arabic}
\include{intro}
\include{intro-aodv}
\include{intro-sq-aodv}
\include{sim-results}
\include{conclusion}
\appendix
\include{app1}
\include{app2}
\bibliographystyle{ieeetr}
\cleardoublepage
\phantomsection
\addcontentsline{toc}{chapter}{References}
\bibliography{Thesis_final}
\end{document}

%% file: ack.tex
\newpage
\thispagestyle{empty}

\hfill \\

\begin{center}
\Large \bf Acknowledgments
\end{center}

With deep sense of gratitude, I would like to thank my adviser, Prof. Abhay Karandikar for his timely input and guidance during the entire course of the project.

My sincere thanks to my co-adviser Dr. Vishal Sharma, Principal Consultant \& Technologist, Metanoia Inc., USA., without whom this thesis would not have been possible. It is a great learning experience to work under his guidance. The regular inputs and encouraging feedbacks have always helped me in doing research in the right direction.

I am very thankful to Hemant Rath for his continuous interaction in the lab during the entire project. I am also thankful to, Punit Rathod for his guidance in scripting and Ashutosh Gore for his valuable inputs and guidance in report writing.

I would like to thank all my fellow labmates for their cooperation in creating a vibrant environment in the lab to work.

I would like to take this as an opportunity to thank Defence Research and Development Organisation for giving me an opportunity to persue my Master's at Indian Institute of Technology, Bombay

I thank my parents for their continuous support and encouragement during my entire Master's course.

\vspace{0.2in}

\begin{flushright} 

{\bf{Mallapur Veerayya}}

{\bf{July 2008}}

\end{flushright}

%% file: abs.tex
\thispagestyle{empty}
\begin{abstract}
An ad-hoc wireless network is a collection of nodes that come together to dynamically create a network, with no fixed infrastructure or centralized administration. An ad-hoc network is characterized by energy constrained nodes, bandwidth constrained links and dynamic topology. With the growing use of wireless networks (including ad-hoc networks) for real-time applications, such as voice, video, and real-time data, the need for Quality of Service (QoS) guarantees in terms of delay, bandwidth, and packet loss is becoming increasingly important. Providing QoS in ad-hoc networks is a challenging task because of dynamic nature of network topology and imprecise state information. Hence, it is important to have a dynamic routing protocol with fast re-routing capability, which also provides stable route during the life-time of the flows.

In this thesis, we have proposed a novel, energy aware, stable routing protocol named, Stability-based QoS-capable Ad-hoc On-demand Distance Vector (SQ-AODV), which is an enhancement of the well-known Ad-hoc On-demand Distance Vector (AODV) routing protocol for ad-hoc wireless networks. SQ-AODV utilizes a cross-layer design approach in which information about the residual energy of a node is used for route selection and maintenance. An important feature of SQ-AODV protocol is that it uses only local information and requires no additional communication or co-operation between the network nodes. SQ-AODV possesses a make-before-break re-routing capability that enables near-zero packet drops and is compatible with the basic AODV data formats and operation, making it easy to adopt in ad-hoc networks.

We demonstrate, through extensive simulation results in NS-2, that the increased route stability afforded by SQ-AODV leads to substantially better QoS performance. Our results show that under a variety of applicable network loads and settings, SQ-AODV achieves packet delivery ratio, on average, 10-15\% better than either AODV or Minimum Drain Rate (MDR) routing protocol, and node expiration times 10-50\% better than either AODV or MDR, with packet delay and control overhead comparable to that of AODV. 
\end{abstract}
\pagebreak

%% file: intro.tex
\chapter{Introduction}
\label{intro}
\section{Background}
\begin{figure}[htbp]
	\centering
	\includegraphics[scale=0.40]{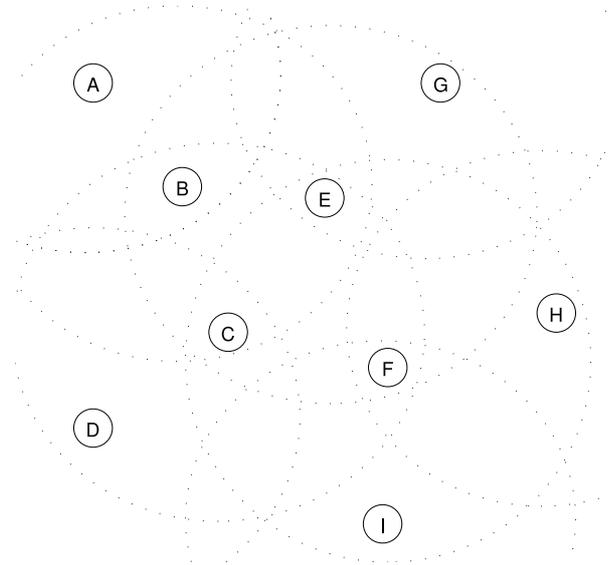}
	\caption{A Simple Ad-Hoc Network with 9 Participating Nodes}
	\label{fig11}
\end{figure}

An ad-hoc wireless network is a collection of nodes that come together
to dynamically create a network, with no fixed infrastructure or centralized
administration as shown in Figure~\ref{fig11}. For a source to send data packets to a destination that is not in its direct range of transmission,
the packets must be relayed through one or more intermediate nodes.
For example, in Figure~\ref{fig11}, if node A wishes to communicate with
node F that is outside of A's direct transmission range, the packets will
have to be relayed either through nodes B and E or through nodes B and C. Hence, a node must act both as a host and a router. A routing protocol is,
therefore, required to find the best possible route to relay a packet
to its desired destination. Two key functions of such a routing protocol are:

\begin{itemize}
 \item Determination of routes for various source-destination pairs
 \item Delivery of data packets to their correct destination
\end{itemize}

Even if the nodes in the ad-hoc network are stationary, the quality of the wireless links between them varies -- both due to the varying amounts of interference created by the transmissions in the network, and due to the variability of the wireless link. Thus, a dynamic routing protocol (as opposed to a set of static routes) is required to find the best possible route to relay a packet to its desired destination.

Ad-hoc networks have a number of applications today, due to their ability to provide an instant network infrastructure to support communications in temporary or mobile environments. For instance, an ad-hoc network is ideal for a battlefield scenario to form a command, control, and communications network for tactical military communications. Another example is the ability to establish a commercial and educational use network in remote areas, where traditional communication infrastructure is non-existent, infeasible or expensive. Since ad-hoc networks can be setup on-demand, with no constraint on connectivity/topology, they also offer unique benefits for applications such as city surveillance networks or networks for law enforcement or rescue/disaster management. In each of these applications, the network in question may be static or semi-static, with different types of data riding on it.

With the growing use of wireless networks (including ad-hoc networks) for real-time applications, such as voice, video, and real-time data, the need for Quality of Service (QoS) guarantees in terms of delay, bandwidth, and packet loss is becoming increasingly important. This is particularly challenging for ad-hoc networks, where the nodes are invariably constrained by energy. Moreover mobility and the time-varying shared wireless medium makes QoS provisioning much more difficult. A key to enabling QoS guarantees in ad-hoc networks, therefore, is a dynamic routing protocol that can adapt quickly to network changes.

Existing routing protocols for ad-hoc networks may be broadly classified into: {\it table-driven}~\cite{dsdv}~\cite{cgsr} and {\it on-demand}~\cite{aodv}~\cite{dsr} protocols. Table-driven protocols are proactive, i.e., they try to keep up-to-date routing information about the entire network through periodic update messages, and can incur significant overhead in many cases. The on-demand protocols, on the other hand, are reactive, i.e., they discover routes as and when required by the sources. Since resources in ad-hoc networks are often scarce, on-demand protocols appear to be more suitable for such networks.

\section{Related Work}
\label{relatedwork}
Standard QoS architectures proposed for the Internet, such as the Integrated Services (IntServ) model~\cite{internet-qos} or the Differentiated Services (DiffServ) model~\cite{internet-qos} are not directly applicable to ad-hoc networks, because they were not designed with the wireless environment in mind. Given the growing importance of QoS in wireless networks, over the last few years, a number of works have proposed ways to improve QoS in an ad-hoc wireless environment.

In~\cite{onqos}, the authors proposed an extension to Ad-Hoc On-demand Distance Vector (AODV)~\cite{aodv} routing protocol to support QoS, assuming the availability of some stationary links in the network. The authors introduced the notion of {\it node stability}, based on a node's history, which incorporated both, a node's mobility and its packet processing ratio. Only stable nodes were considered for routing. However, the authors did not consider the impact that unpredictable link failures would have on re-routing.

In~\cite{quasar} the authors proposed a Quality-of-Service aware Source initiated Ad-hoc Routing (QuaSAR) protocol, which adds QoS control mechanisms to all phases of source routing. The protocol maintains an estimate of the
battery power required by the application, and uses that in the path selection process, while attempting to give guarantees on the latency and bandwidth of a flow. The route is selected by the source, after collecting statistics on all possible routes that satisfy a flows latency, bandwidth, and power requirements. Energy consumption in routing a flow is minimized by choosing
the ``shortest" path, by mapping signal strength to distance. A characteristic of QuaSAR is that it effectively trades reduced packet drops for increased protocol overhead. The route request messages in QuaSAR must carry latency,
bandwidth, signal strength, and battery power information for all nodes along the path to enable the selection of a path that satisfies the QoS requirements of a session. QuaSAR requires deploying a completely new protocol, and suffers
from poor scalability, since source routing can cause the length of route request and route reply messages to become excessive in larger networks.
Furthermore, the basic QoS capabilities of QuaSAR are examined for a rather limited scenario, with only a single source and single (mobile) destination.

In~\cite{wt-based-stable} authors have proposed a stable, weight-based, on-demand routing protocol. The difference with QuaSAR is that instead of carrying the different parameters themselves, the authors use them to compute a composite ``weight", which is the one carried in the protocol messages. The weight used to select stable routes is based on three components: Route Expiration Time (RET), which is the predicted time of link breakage between two nodes due to mobility, Error Count (EC), which captures the number of link failures due to mobility, and Hop Count (HC). The authors have assumed that all nodes are synchronized via a Global Positioning System (GPS), so that
two adjacent nodes may predict the RET. While the proposed scheme may combat against link breaks due to mobility,  link breaks due to the draining node energy is a factor that also must be accounted for when computing weights for stable routing.

In~\cite{multiconstraint-qos}, the authors have proposed a stable route selection scheme based on Link Expiration Time Threshold ($LET_{th}$). The Link Expiration Time ($LET$) is computed based on a prediction of neighbor mobility. $LET$ computation needs to know the position of the neighbors, and hence requires periodic topology updates. However, the authors have not considered the impact that unpredictable link failures would have on re-routing.

In~\cite{mdr}, the authors proposed a new metric, Energy-Drain-Rate (EDR),  which is defined as the rate at which energy is consumed at a given node at time $t$. The  corresponding cost function is defined as:
\begin{equation*}
C_{r} = min T_{r}^{i}{(t)},
\end{equation*}
where,
\begin{equation*}
T_{r}^{i}{(t)} = \frac{E_{r}^{i}{(t)}}{DR_{i}{(t)}}, 
\end{equation*}
where, $DR_{i}{(t)}$ is the drain rate of node $i$ at time $t$ and $E_{r}^{i}{(t)}$ is the residual battery power of node $i$ at time $t$.

Thus the life-time of a path is determined by the minimum $T_{r}^{i}{(t)}$ along that path. The Minimum Drain Rate (MDR)~\cite{mdr} mechanism selects the route with maximum life-time. Each node monitors its energy consumption during a given past interval $\tau$ and maintains the drain rate value using an exponential weighted moving average. The proposed MDR algorithm attempts to select the best possible stable route for a given source and destination. The periodic 
route update used in MDR, however, soon becomes costly, as it increases control overhead and degrades performance at higher network loads.

From the proposals reviewed so far~\cite{onqos},~\cite{wt-based-stable},~\cite{multiconstraint-qos},~\cite{mdr} it is clear that there is a need for a stable routing protocol that can provide stability to the routes selected for routing QoS-enabled applications, and also has mechanisms for fast re-routing to tackle unpredictable link
breakages. Furthermore, for the scheme to be scalable, the stability should come at minimum or no overhead.

\section{Contributions of the Thesis}
\label{contributions}
A key to providing QoS guarantees in ad-hoc networks is to find, not just any route to the desired destination, but rather a route that can, with high probability, survive for the duration of the session. This ensures that communication once initiated will not be disturbed. It is also useful to have a mechanism to quickly find an alternate route, if one exists, for the session, in case of unpredictable link failure.

In this thesis, we have proposed an energy-aware on-demand routing protocol which also provides stable routes to the flows during their life-time to support QoS in ad-hoc networks. The proposed energy-aware, stable routing protocol named, Stability-based QoS-capable Ad-hoc On-demand Distance Vector (SQ-AODV) protocol is an enhancement of the well-known AODV~\cite{aodv} protocol for ad-hoc wireless networks. SQ-AODV utilizes a cross-layer design approach in which information about the current residual energy, average energy drain-rate of a node, and the session-duration (if known) of the application is used to find a stable route. SQ-AODV also does a proactive route maintenance using \textit{``make-before-break"} mechanism for finding an alternate route for the session when the energy drain rate of a node suggests that it will cease forwarding before the session is completed. Since SQ-AODV uses only local information and requires no additional communication or co-operation between the network nodes, it increases the packet delivery ratio in the network at \textit{virtually no overhead} making it more suitable for ad-hoc wireless environment.

\section{Organization of the Thesis}
\label{organization}
The thesis consists of 5 chapters. Chapter~\ref{intro}, gives a brief introduction to wireless ad-hoc networks, and need for QoS support and dynamic routing protocol in wireless ad-hoc networks. Some of the proposed QoS and stability-based routing protocols are also reviewed in this chapter. Chapter~\ref{intro-aodv}, gives a brief introduction to AODV routing protocol, which we have modified to realize SQ-AODV in NS-2~\cite{NS-2}. A detailed description and operation of SQ-AODV is presented in Chapter~\ref{intro-sq-aodv}. In Chapter~\ref{simulations}, we have given a complete details of our simulation scenarios, results along with discussions. Finally, Chapter~\ref{conclusion} give the summary of the thesis along with some of the future works.

%% file: intro-aodv.tex
\chapter{Ad-Hoc On-demand Distance Vector (AODV) Routing Protocol}
\label{intro-aodv}
In this chapter, we give a very brief introduction to AODV~\cite{aodv}, which we have modified in designing SQ-AODV. Since the phases of operation of SQ-AODV remain the same as those of AODV, it is important to first understand the basic operation of AODV. AODV is a destination-based reactive protocol. It avoids routing loops by tagging an unique sequence number to route information for each destination. This sequence number is generated or originated by the destination. AODV for its operation assumes symmetric links between neighboring nodes. That is, the links are bidirectional, and should have same properties in both directions. AODV routing protocol uses different routing messages to discover the routes and maintain links.

\begin{itemize}
 \item RREQ is a route request message used whenever a new route to a
 destination is required.
 \item RREP is a reply message for a route request.
 \item Periodic HELLO messages are broadcast to check the presence
 of immediate active neighbors.
\end{itemize}

If a node does not lie along an active route, it neither
maintains routing information nor participates in the exchange of routing
information. The protocol consists of two basic processes:

\begin{enumerate}
 \item Path discovery
 \item Path maintenance
\end{enumerate}

\section{Path Discovery}
\label{aodv-path-dis}
A path discovery process is initiated whenever a source node needs to send
data packets to a destination node and does not have a route information for
this destination node. Consider the network in Figure~\ref{fig11}. Suppose
node A wants to send data packets to node G and does not have a route
information. Then node A initiates a route discovery process by broadcasting
a RREQ packet to its immediate neighbors (in our case node B).
Each intermediate node after receiving the first RREQ packet does the following:

\begin{itemize}
 \item Checks whether it has a current route information about the
 destination node
 \item If it has a current route information, it sends a RREP back to the source node
 \item If it does not have a current route information, it rebroadcasts the RREQ
 to its neighboring nodes and keeps a record of the following information for
 setting up a reverse path:
	\begin{itemize}
 	\item Destination IP address
 	\item Source IP address
 	\item Broadcast ID
 	\item Expiration time for the reverse path entry
 	\item Source node sequence number
\end{itemize}
\end{itemize}

Every subsequent RREQ (copies) with the same broadcast ID is discarded.
In our case assume node B rebroadcasts the RREQ to its immediate active
neighbors A, C and E and keeps a record. Node C after receiving RREQ
rebroadcasts the RREQ to its immediate neighbors B, D, E and F, and
keeps a record. Node D simply times out because its only neighbor
is node C from which it has received RREQ. All intermediate nodes repeat
this operation of either rebroadcasting or timing out till RREQ reaches the final destination. Finally, when RREQ reaches the desired destination
node G, the node will unicast a RREP message back to the source through the
\textit{reverse path setup}.

\subsection{Reverse Path Setup}

\begin{figure}[htbp]
 \centering
 \includegraphics[scale=0.40]{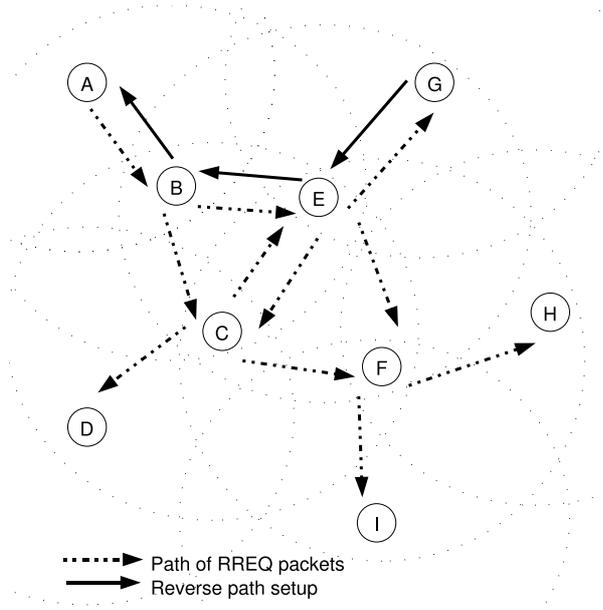}
 \caption{Reverse Path Setup in AODV}
 \label{fig21}
\end{figure}

A source sequence number is used to maintain freshness information
about the reverse path to the source.
In Figure~\ref{fig21}, RREQ travels form node A to various active
intermediate nodes and when it finally reaches the destination node G,
it automatically sets up the reverse path from the destination to source.
To do this reverse path setup every intermediate node (in our case nodes B
and E) records the address of the active neighbor from which it received the
first copy of the RREQ. These reverse path entries are maintained for
sufficient amount of time so that the RREQ packet traverses the network
and produce a reply back to the source. The reverse path that is setup
from node E to node A is indicated by solid arrows in Figure~\ref{fig21}.

\subsection{Forward Path Setup}

\begin{figure}[htbp]
 \centering
 \includegraphics[scale=0.40]{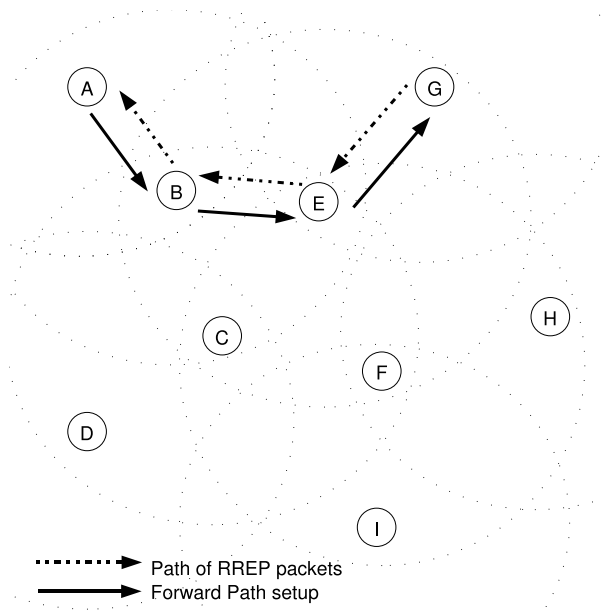}
 \caption{Forward Path Setup in AODV}
 \label{fig22}
\end{figure}

As the RREP travels back to the source using a reverse route, either from the destination node or an
intermediate node that has a current route information about the destination, each
node along this reverse path sets up a forward pointer to the node from
which the RREP is received. Each node also updates the timeout information
for this source to destination, and records the latest destination sequence
number. In Figure~\ref{fig22} solid arrows indicate the forward path from
node A to node G. This path is setup with the help of forward pointers
as the RREP travels back from destination node G to source node A.
The nodes that are not active along the path determined by the RREP will timeout and delete the reverse pointers. Once the forward path is
setup and the RREP reaches the source node A, source node A will immediately
start data transmission.

\subsection{Route Table Management}
Apart from the source and destination sequence numbers as entries in the routing
table, there are expiration timers associated with reverse path entries and
route invalidation. The purpose of the timer meant for reverse path entry
is to give timeout information for purging of those reverse path entries
in the nodes that do not lie along the path determined by RREP. In
Figure~\ref{fig21} nodes C, D, F, H and I purge there entries after this
timer expiration. This expiration time depends on the size of the ad-hoc network.
There is another expiration timer which is used to invalidate a route
already available in the cache. A neighbor is considered to be active,
if it originates or relays at least one packet for that destination in
the timeout period and the address of the active neighbors is also entered
in the table. Hence a node maintains the fallowing information for each route
table entry:

\begin{itemize}
 \item Destination IP address
 \item Next hop
 \item Number of hops
 \item Sequence number for the destination
 \item Active neighbors for that route
 \item Expiration time for the route table entry
\end{itemize}

If there is more than one route entry for a particular destination, the
node chooses the one with higher sequence number. If the sequence
numbers are same then a route with smaller metric (less number of hop count)
is chosen.

\section{Path Maintenance}
\label{aodv-path-main}
If a node moves from the current location to a new location in the network,
the routing will not be affected unless this node was in the active
routing path. When a source node moves to a new location in the network and affects the route of an active session (i.e., before route invalidation timer expires), then the source can re-initiate
a route discovery procedure if a route to the desired destination is still required.
On the other hand, if an intermediate node of an active session moves from its present position, then a
special RREP is sent to the affected source nodes. Periodic HELLO
messages are used to detect link failures. Link failures can
also be detected if a node is unable to forward a packet to the next hop.
Once a link failure is detected, an unconditional RREP with fresh
sequence number and hop count set to infinity is broadcast to active
neighbors. Within some time all active nodes in the network will
know about link failure. The source nodes can restart the discovery
process if they still need a route to a destination.

\subsection{Local Connectivity Management}
Nodes learn about their neighbors in two ways. One way is whenever a
node receives a broadcast message from a neighbor it updates its local
connectivity. Other way is to broadcast HELLO messages to its active
neighbors. If a node does not receive HELLO messages consecutively,
this indicates that local connectivity is changed.

%% file: intro-sq-aodv.tex
\chapter{Stability-based QoS-capable AODV (SQ-AODV)}
\label{intro-sq-aodv}
In this chapter, we present our energy-aware on-demand routing protocol SQ-AODV. We explain the operation of SQ-AODV along with the cross-layer design used to implement the protocol in NS-2~\cite{NS-2}.

\section{Introduction}
SQ-AODV is an enhancement to the well-known AODV~\cite{aodv} routing protocol, which we have discussed in Chapter~\ref{intro-aodv}. The enhancements are done in both \textit{Path Discovery} and \textit{Path Maintenance} phases of AODV to make it a stable and dynamic routing protocol. The two main features of SQ-AODV are it:

\begin{itemize}
\item Provides stable routes by accounting for the residual life-time at intermediate nodes (calculated using the current Average-Energy-Drain-Rate (AEDR)) and the duration of the session (if known) at the route selection stage.
\item Guards against link breakages that arise when the energy of a node(s) along a path is depleted, by performing a make-before-break re-route (where possible). This minimizes packet loss and session disruptions.
\end{itemize}

The first feature ensures that SQ-AODV only routes sessions along routes that either have intermediate nodes with sufficient energy to last the length of the session or along routes that maximize the residual life-time of the bottleneck node, thus ensuring, with very high probability, that session disruption due to energy depletion at an intermediate node does not occur. It turns out that this increased stability leads to substantially better QoS in terms of packet delivery ratio (PDR) and packet delay (PD), even without explicitly accounting for bandwidth, jitter or delay requirements, as our subsequent results demonstrate.

The second feature ensures that when a link break due to node energy depletion is imminent, SQ-AODV proactively re-routes sessions, without losing any packets. Once again, this provides near-zero packet loss and superior QoS performance.

\begin{figure}[ht]
	\centering
	\includegraphics[scale=0.35]{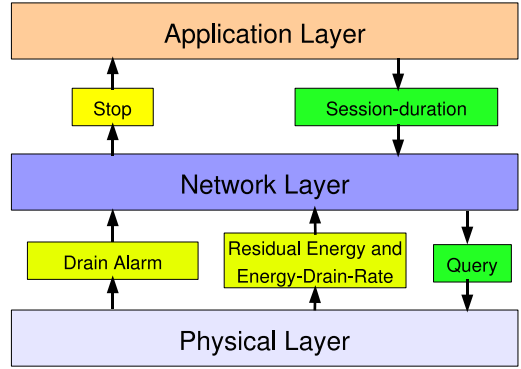}
	\caption{Cross-layer Design Used in SQ-AODV}
	\label{fig31}
\end{figure}

The operation of SQ-AODV utilizes the cross-layer design depicted in Figure~\ref{fig31}, where energy information from the physical layer is used in admission control decisions at the network layer and to turn-off sessions at the application layer. We now explain these two features in more detail.

\subsection{Path Discovery}
The first modification/feature is outlined in Algorithm 1, and helps in choosing an appropriate sequence of intermediate nodes for the requesting session. 

\begin{algorithm}[ht]
\caption{: Selection of an Intermediate Node as Router}
\label{algo31}
\begin{algorithmic}[1]
\STATE An intermediate node \textbf{N} receives a \textbf{RREQ};
\IF {\textit{Session-Duration} is specified in the \textbf{RREQ}}
\STATE Check
	\IF { \textit{Current-Energy} $>$ (\textit{Session-Duration} $\times$ \textit{AEDR}) }
	\STATE Update \textbf{Bottleneck life-time} field of \textbf{RREQ};
	\STATE \textbf{ADMIT} the session and forward the \textbf{RREQ} to the neighbors
	\ELSE
	\STATE \textbf{REJECT} the session, and \textbf{DROP} the \textbf{RREQ}
	\ENDIF
\ELSE
	\IF {\textit{Current-Energy} $>$  \textbf{Threshold-1} \\}
	\STATE Update \textbf{Bottleneck life-time} field of \textbf{RREQ};
	\STATE \textbf{ADMIT} the session and forward the \textbf{RREQ} to the neighbors
	\ELSE
	\STATE \textbf{REJECT} the session, and \textbf{DROP} the \textbf{RREQ}
	\ENDIF
\ENDIF
\end{algorithmic}
\end{algorithm}

The application layer of a source that wishes to communicate with a destination, generates data packets and transmits them to the network layer. At the network layer, the routing protocol responsible for finding a route to the desired destination initiates a route discovery procedure, if it does not already have a route for that destination. We assume here that, if the session-duration is known, the application layer directly provides that
to the network layer, as shown in Figure~\ref{fig31}. If not, each intermediate node uses a heuristic and accepts a session only if it has at least \textbf{Threshold-1} of residual life (\textbf{Threshold-1} is the residual energy of a node with which the node is alive for the next X seconds at current AEDR, in our implementation X = 5 seconds).

The source broadcasts Route Request (RREQ) packets  to its neighbors when it has no route to the desired destination. When a RREQ packet reaches an intermediate node, Algorithm 1 queries the physical layer for the current residual energy, and checks whether the residual energy at the current
AEDR is sufficient to last the duration of the flow. The session is only admitted if that is the case. If the session-duration is unknown, the algorithm admits the session only if the residual energy at the node is above \textbf{Threshold-1}. Before forwarding, the node updates the bottleneck life-time field of the RREQ packet.

The Energy-Drain-Rate (EDR) is computed as a  difference between  
the energy $En$ of the node at periodic intervals divided by the length of the interval. Thus,

\begin{center}
\textit{EDR($t_{2}$)} = $\frac{En(t_{1}) - En(t_{2})}{t_{2}-t_{1}}$,
\end{center}

where $En(t_{1})$ and $En(t_{2})$ are energy levels of the node at times $t_{1}$ and $t_{2}$ respectively. Finally, this EDR is averaged using exponential averaging with $\alpha=0.5$ to compute the AEDR as follows:

\begin{center}
\textit{AEDR(t)} = $\alpha$ $\times$ \textit{EDR(t)} + (1-$\alpha$) $\times$ \textit{AEDR(t-1)}.
\end{center}

Finally, when the RREQ packets reach the destination, it picks a route that maximize the route life-time by selecting the one with maximum life-time of the bottleneck node.

\subsection{Path Maintenance}
The second modification/feature helps the routing protocol to adapt quickly to imminent link breakage likely to occur when the energy of a node is fully drained. The algorithm for this is depicted in Algorithm 2. Since
the physical layer keeps track of the AEDR, it sends an alarm to the network layer shortly before it is about to drain completely i.e., when the current energy of the node is less than \textbf{Threshold-2} (\textbf{Threshold-2} is the residual energy of a node with which the node is alive for the next Y seconds at current AEDR, in our implementation Y = 1 second). The routing protocol adapts to this event, and its behavior depends on whether the node  is an intermediate node (\textbf{I}) or a destination node (\textbf{D}). 

If the node receiving the drain alarm from its physical layer is an
intermediate node, it sends a Route Change Request (RCR) packet  to all source nodes using it as an intermediate hop towards their respective destinations.
The source upon receiving the RCR packet, begins a new route discovery procedure for the session,  and thus, with high probability, finds a new route
before an actual link break occurs on the original route, leading to the \textit{make-before-break} behavior. This reduces packet drops due to link breakage and the consequent delay incurred, and enables the routing protocol to quickly adapt to network changes, if an alternate path to the desired destination exists. If the node being drained is a destination node, it sends a request to the source to stop all traffic transmission to itself. When the request reaches the source, the  network layer sends a stop signal to the application, as shown in Fig.~\ref{fig31}, preventing further transmission of data. This reduces the number of packet drops in the network and increases packet delivery ratio, and reduces resource usage by avoiding packet transmissions to unavailable destinations. If a source node itself is about to drain, it simply continues to transmit data until it cannot transmit anymore.

\begin{algorithm}[ht]
\caption{: Route Maintenance by Make-before-break}
\label{algo32}
\begin{algorithmic}[1]
\STATE Node \textbf{N} periodically compute the \textbf{EDR} and check for \textit{Current-Energy};
	\IF {\textit{Current-Energy} $<$ \textbf{Threshold-2}\\}
	\STATE Check
		\IF {\textbf{N} == \textbf{I} \\}
		\STATE  Send \textbf{RCR} to all the source nodes using this node as router
		\ENDIF
		\IF {\textbf{N} == \textbf{D} \\}
		\STATE Send a \textbf{Stop-Traffic} request to all sources that are communicating with this destination
		\ENDIF
	\ENDIF
\end{algorithmic}
\end{algorithm}

Note that SQ-AODV uses only local information, and requires no additional communication or co-operation between the nodes. Indeed, the algorithms described above could be used with any underlying routing protocol, but we use AODV protocol as it is one of the most popular ad-hoc routing protocols.

%% file: sim-results.tex
\chapter{Simulation Experiments, Results and Discussions}
\label{simulations}

In this chapter, we present and discuss a wide range of results of extensive simulations that we have conducted in NS-2~\cite{NS-2} (Version 2.30), to compare the performance of SQ-AODV with that of MDR~\cite{mdr} and AODV~\cite{aodv}. We have considered the following six parameters:

\begin{itemize}
\item \textbf{Packet Delivery Ratio (PDR)}: is the ratio of the number of packets successfully received
by all destinations to the total number of packets injected into the network by all sources. The PDR is
therefore a number between 0 and 1.

\item \textbf{Normalized Control Overhead:} is the ratio of number of routing packets transmitted 
(hop wise) by all the nodes to the total number of packets successfully received by all destinations in the 
network. The normalized control overhead is therefore a number greater than 0.

\item \textbf{Average Packet Delay:} is the sum of the times taken by the successful data packets to 
travel from their sources to destinations divided by the total number of successful packets. The average packet delay 
is measured in seconds.

\item \textbf{Average Hop Count:} is the sum of the number of hops taken by the successful data packets to 
travel from their sources to destinations divided by the total number of successful packets. The average hop count  
is measured in number of hops.

\item \textbf{Node Expiration Time (NET)}: is the time for which a node has been alive before it must halt transmission due to battery depletion. The node expiration time is plotted as number of nodes alive at a given time, for different points in time during the simulation.

\item \textbf{Connection Expiration Time (CET)}: is the time for which a connection has been active before it must cease transmission due to the non-availability of a route between source and destination.
This occurs when nodes along the path expire or become unreachable due to poor link conditions. The connection expiration time is expressed in seconds.
\end{itemize}

\noindent
We present our simulation results in 3 different parts, and are as follows:

\begin{enumerate}
\item Demonstration of SQ-AODV features
\item Validation of MDR~\cite{mdr} implementation
\item Performance comparison of SQ-AODV, MDR and AODV~\cite{aodv}
\end{enumerate}

\section{Demonstration of SQ-AODV Features}
\label{sq-aodv features}

In this section, we present our simulation results to demonstrate the energy awareness and make-before-break re-routing features of SQ-AODV by comparing the packet delivery ratio performance with that of AODV.

\subsection{Simulation Setup}
\begin{figure}[htbp]
	\centering
	\includegraphics[scale=0.40]{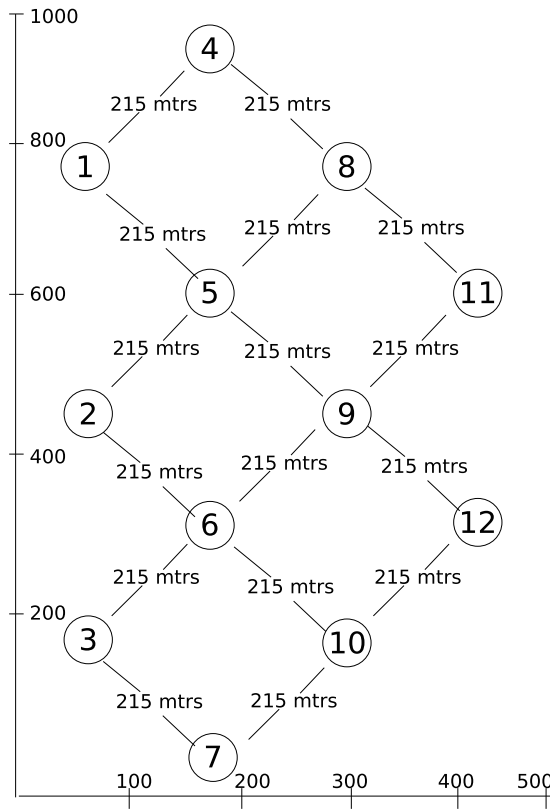}
	\caption{12 Node Topology}
	\label{fig51}
\end{figure}

We have considered the 12-node topology in 500 m x 1000 m area as shown in Figure~\ref{fig51}. The nodes are identical in their capability, but are initialized with different energies in different experiments Expt1 to Expt5, that we have conducted, and there is no mobility in the network.

To calibrate the load that can be supported by the network, an extensive series of simulations with one, two, three, and five simultaneous sessions was conducted with varying average session data rates. We found that when the aggregate rate of sessions at the nodes exceeded about 225 Kbps on average, the packet drop rate in the network became excessive, indicating that the network was saturated beyond capacity. We have thus used this rate as 100\% load, and normalized using this.

The MAC layer protocol is IEEE 802.11 DCF  (Distributed Co-ordination Function) with the PLCP (Physical Layer Convergence Protocol)  data rate being 1 Mbps. The parameters used in the simulations for Expt1 to Expt5 are listed in Table~\ref{tab51}.

\begin{table}[htbp]
\centering
  \caption{Values of Parameters Used for the Simulation of the 12 Node Topology}
  \hfill \\	
  \begin{tabular}{|c|c|}
  \hline
  Packet size & 512 Bytes \\
  \hline
  Packets/Session & 1000 \\
  \hline
  Date traffic & Poisson with exponential\\
  \            & inter-arrival time \\
  \hline
  MAC Protocol & IEEE 802.11 DCF \\
  \hline
  PCLP Data rate & 1 Mbps \\
  \hline
  Buffer length & 50 Packets \\
  \hline
  Transmit power & 0.2818 W \\
  \hline
  Propagation model & Two-Ray Ground \\
  \hline
  \end{tabular}
  \label{tab51}
\end{table}

Here we demonstrate the PDR performance of SQ-AODV under a variety of different network loads and node energies, and assess the benefit of its make-before-break strategy. In each experiment Expt1 to Expt5, the initial energy levels of the nodes is randomly chosen between 10 Joules to 100 Joules, and compared the performance of PDR for SQ-AODV and AODV. The starting times of the sessions were chosen such that there were at most between 3-5 sessions in parallel in the network at any instant. The network load was varied from about 20\% to 90\%, so the session data rate is varied from 15 Kbps to 65 Kbps (3-5 sessions of 15 Kbps each equates to 20\% of 225 Kbps (225 Kbps equals 100\% load) and 3-5 sessions of 65 Kbps each equates to 90\% of 225 Kbps). Each experiment is run 50 times (each initialized with different seed), and the resulting PDR is averaged over these 50 runs. There are 12 sessions in each experiment, with 1000 packets per session, generated as per a Poisson process. The energy distribution of the nodes for each of the 5 experiments Expt1 to Expt5, as well as the source-destination pairs for each session are summarized in Table~\ref{tab52} and Table~\ref{tab53}, respectively.

\begin{table}[htbp]
\centering
  \caption{Initial Energy of Nodes for Each Experiments in Joules}
  \hfill \\	
  \begin{tabular}{|c|c|c|c|c|c|}
  \hline
  Node & Expt1 & Expt2 & Expt3 & Expt4 & Expt5 \\
  \hline
  $1$ & $45$ & $29$ & $94$ & $82$ & $93$ \\
  \hline
  $2$ & $20$ & $91$ & $40$ & $73$ & $39$ \\
  \hline
  $3$ & $24$ & $40$ & $29$ & $37$ & $64$ \\
  \hline
  $4$ & $62$ & $97$ & $87$ & $33$ & $31$ \\
  \hline
  $5$ & $59$ & $82$ & $73$ & $90$ & $73$ \\
  \hline
  $6$ & $93$ & $82$ & $50$ & $30$ & $35$ \\
  \hline
  $7$ & $85$ & $50$ & $87$ & $64$ & $100$ \\
  \hline
  $8$ & $68$ & $34$ & $54$ & $28$ & $19$ \\
  \hline
  $9$ & $24$ & $15$ & $24$ & $86$ & $61$ \\
  \hline
  $10$ & $30$ & $17$ & $18$ & $86$ & $49$ \\
  \hline
  $11$ & $55$ & $43$ & $35$ & $90$ & $87$ \\
  \hline
  $12$ & $39$ & $20$ & $95$ & $11$ & $97$ \\
  \hline
  \end{tabular}
  \label{tab52}
\end{table}

\begin{table}[htbp]
\centering
  \caption{Source-destination Pairs in the 12 Node Topology}
  \hfill \\	
  \begin{tabular}{|c|c|c|c|}
  \hline
  $Session\ No.$ & $\{src,\ dst\}$\\
  \hline
  $1$ & $\{1,\ 11\}$\\
  \hline
  $2$ & $\{2,\ 12\}$\\
  \hline
  $3$ & $\{3,\ 11\}$\\
  \hline
  $4$ & $\{8,\ 6\}$\\
  \hline
  $5$ & $\{10,\ 1\}$\\
  \hline
  $6$ & $\{2,\ 4\}$\\
  \hline
  $7$ & $\{12,\ 3\}$\\
  \hline
  $8$ & $\{4,\ 2\}$\\
  \hline
  $9$ & $\{11,\ 1\}$\\
  \hline
  $10$ & $\{7,\ 6\}$\\
  \hline
  $11$ & $\{8,\ 10\}$\\
  \hline
  $12$ & $\{12,\ 1\}$\\
  \hline
  \end{tabular}
  \label{tab53}
\end{table}

\subsection{Results and Discussions}

\begin{figure}[htbp]
	\centering
	\includegraphics[scale=0.75]{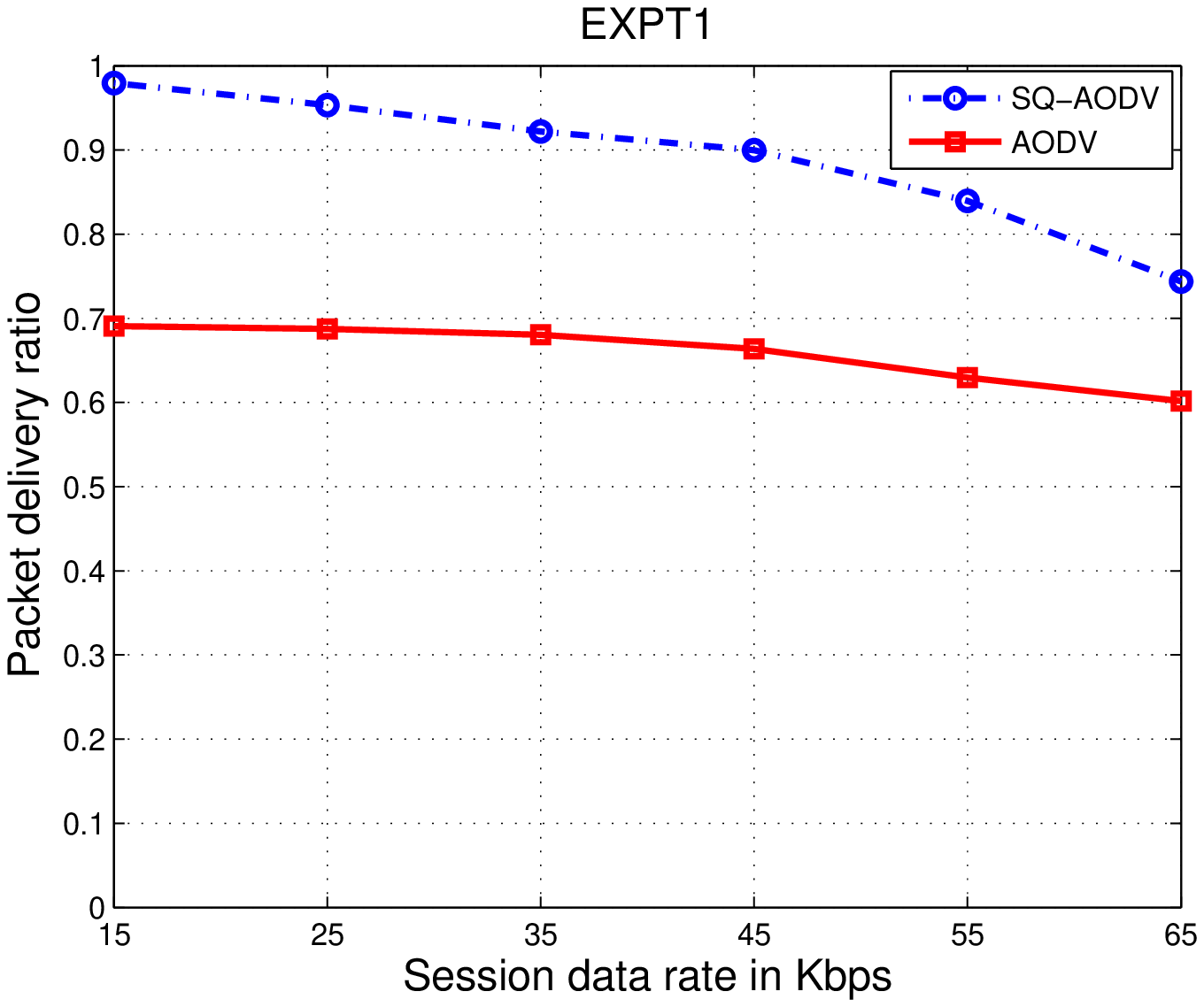}
	\caption{Packet Delivery Ratio for EXPT1}
	\label{fig52}
\end{figure}

\begin{figure}[htbp]
	\centering
	\includegraphics[scale=0.75]{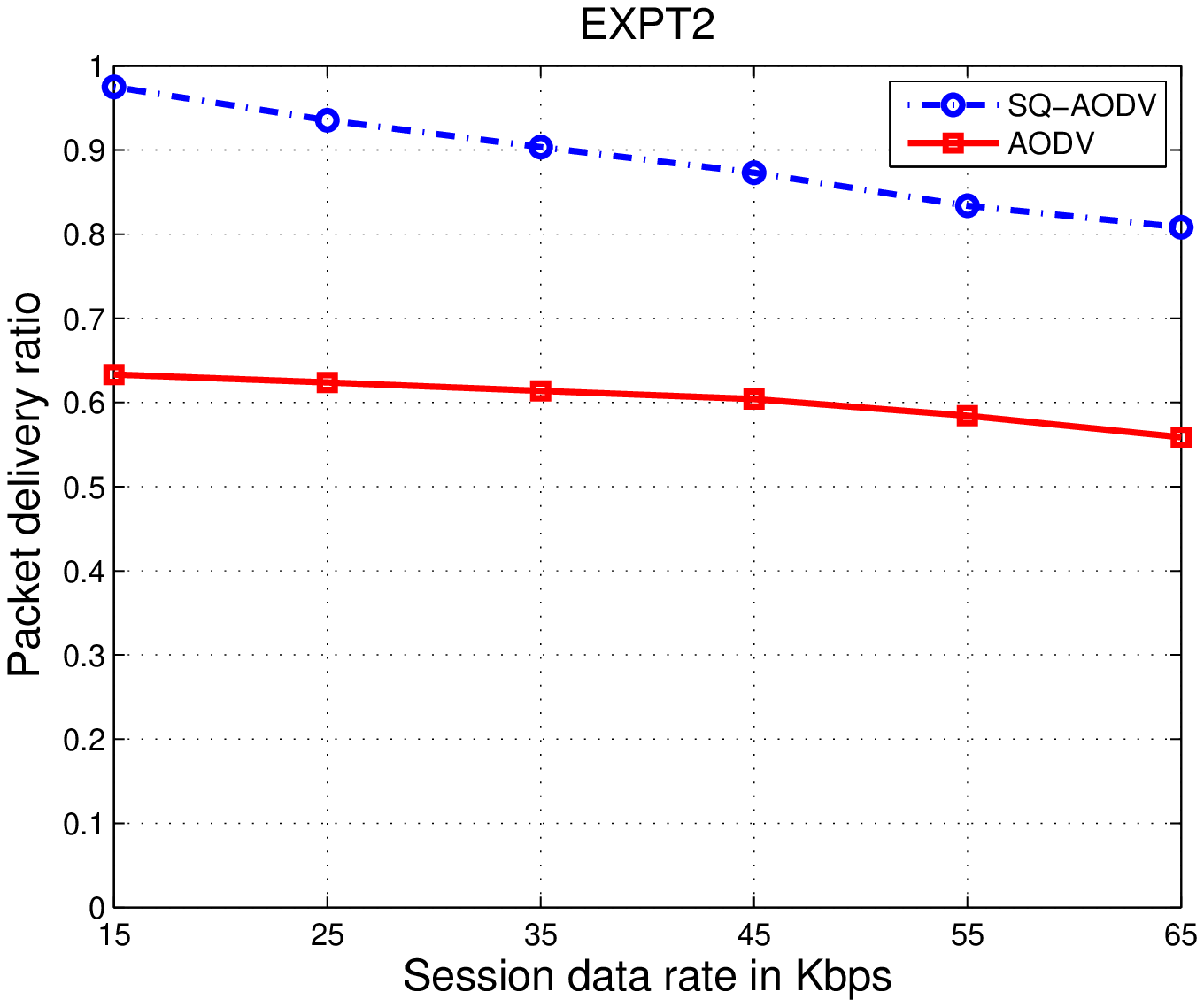}
	\caption{Packet Delivery Ratio for EXPT2}
	\label{fig53}
\end{figure}

\begin{figure}[htbp]
	\centering
	\includegraphics[scale=0.75]{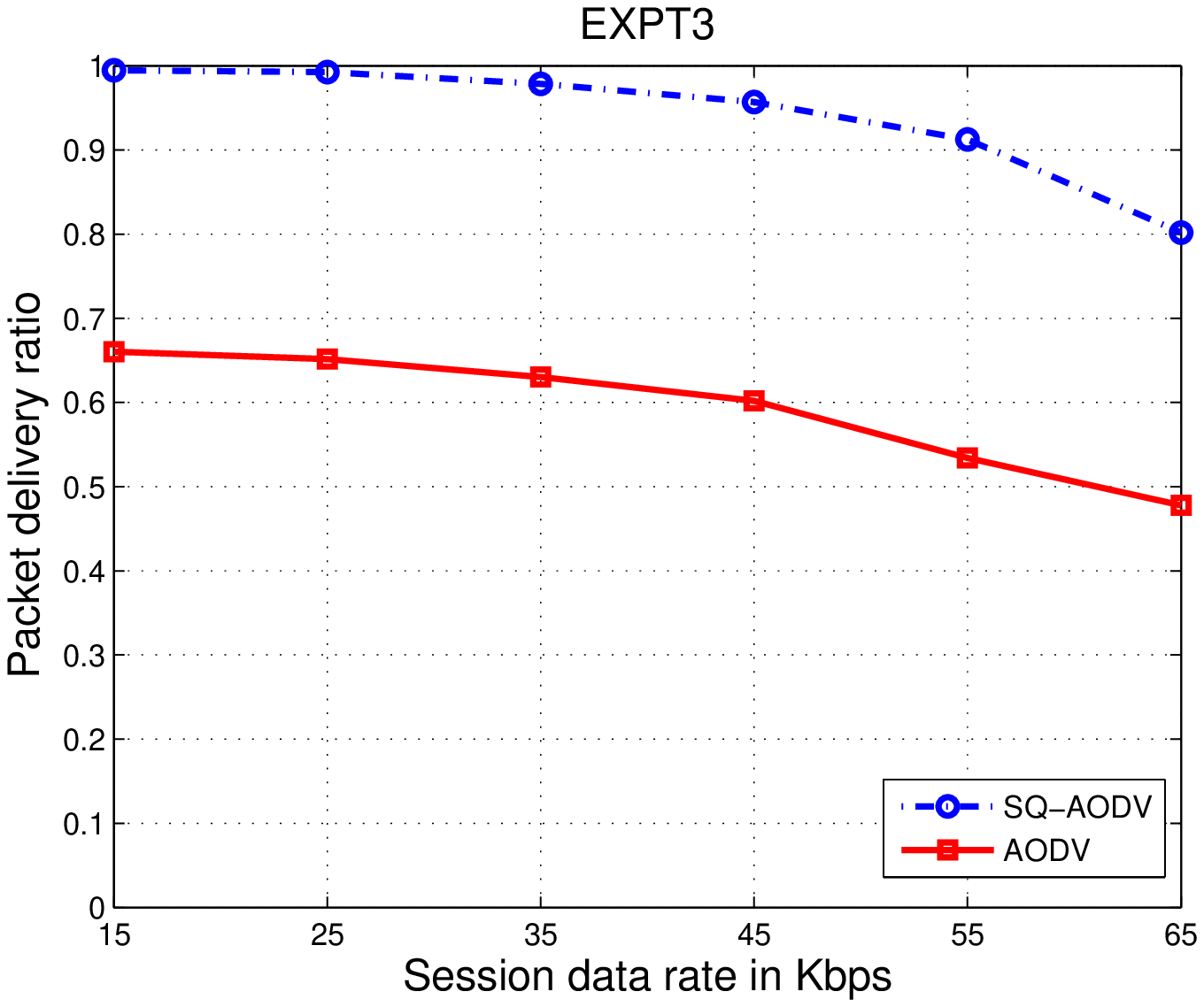}
	\caption{Packet Delivery Ratio for EXPT3}
	\label{fig54}
\end{figure}

\begin{figure}[htbp]
	\centering
	\includegraphics[scale=0.75]{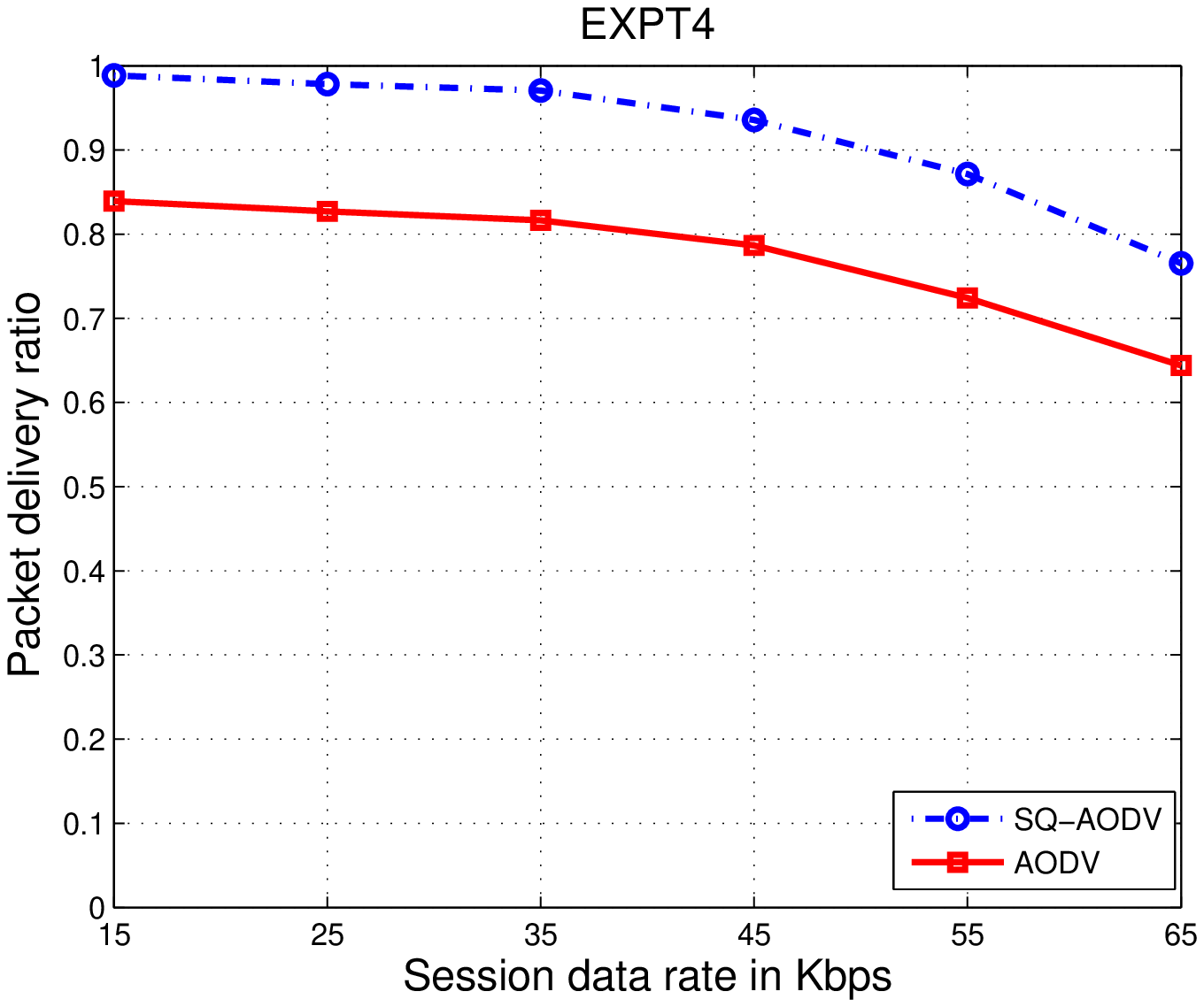}
	\caption{Packet Delivery Ratio for EXPT4}
	\label{fig55}
\end{figure}

\begin{figure}[htbp]
	\centering
	\includegraphics[scale=0.75]{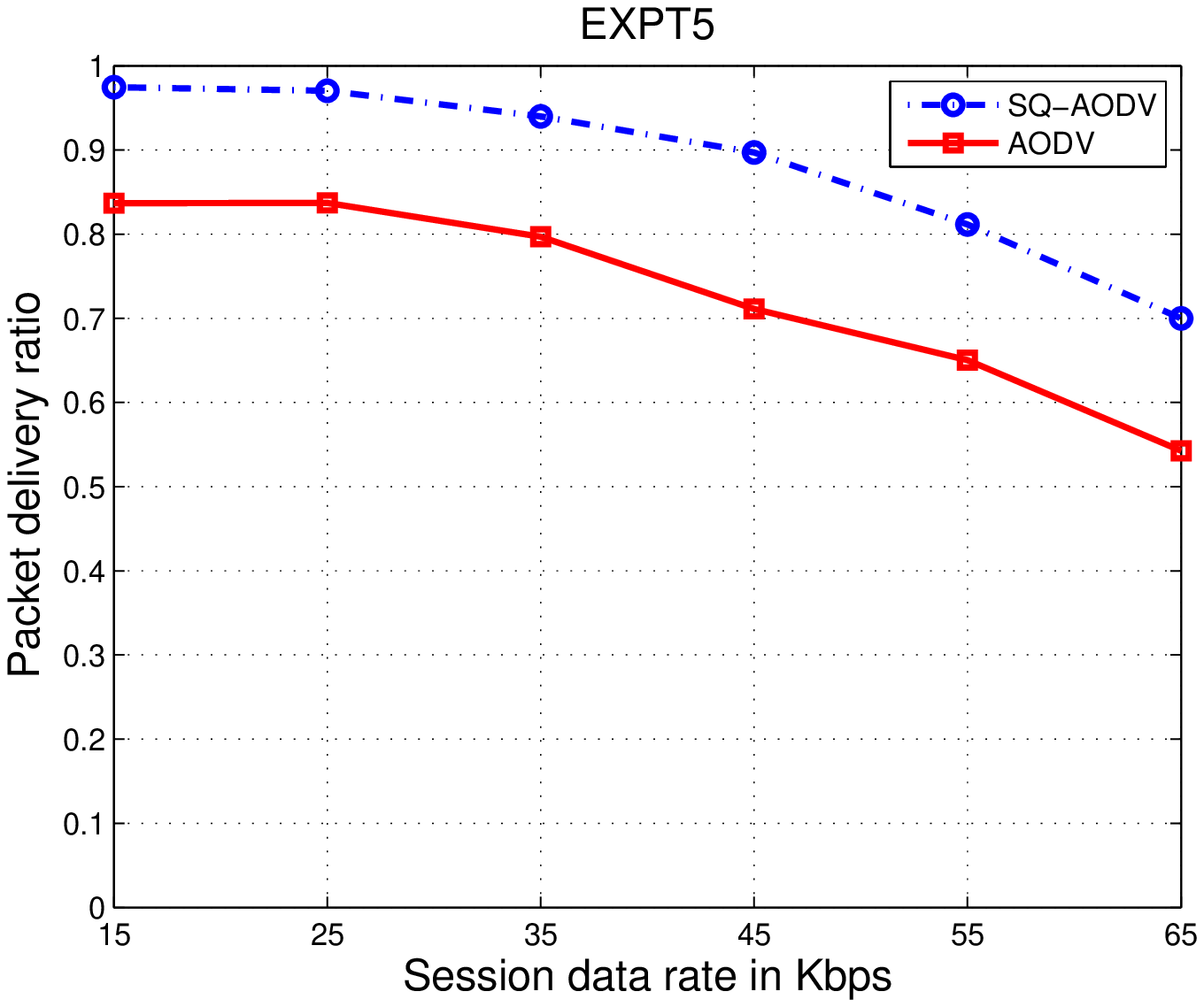}
	\caption{Packet Delivery Ratio for EXPT5}
	\label{fig56}
\end{figure}

The packet delivery ratio for each of the experiments Expt1 to Expt5, for both SQ-AODV and AODV~\cite{aodv} is plotted in Figures~\ref{fig52} - \ref{fig56}. It can be observed that, in all cases, and for all loads, the packet delivery ratio is improved substantially, and SQ-AODV outperforms AODV. This is because there are some nodes whose battery life is limited. Choosing these nodes as intermediate nodes, as is done by AODV, leads to disconnections in the session. SQ-AODV, on the other hand, performas better because, (i) it chooses those intermediate nodes whose energy is sufficient to support the session for its entire durations or it chooses those intermediate nodes which maximize the life-time of the route, and (ii) its make-before-break strategy that re-routes a session proactively when a  link failure due to depletion of node energy is imminent. Thus, SQ-AODV is successful in reducing packet drops in the network drastically. Additionally, in SQ-AODV, the traffic of a source is stopped just as a destination is about to drain, leading to saving of network resources, by not transmitting packets that would, in any case, not be used by the destination.

\section{Validation of MDR Implementation}
\label{MDR-verification}
In this section we give a brief introduction to the Minimum Drain Rate (MDR)~\cite{mdr} routing protocol and its implementation in NS-2~\cite{NS-2}. We also present the simulation results to demonstrate the correctness of our implementation.
Since MDR is an energy-aware routing protocol with a system model and operation quite similar to that of SQ-AODV, we have chosen MDR for comparison with SQ-AODV.

\subsection{MDR Implementation}
\label{mdr-implementation}
\textit{Minimum drain rate} is basically a mechanism used to select a path between a source and destination that maximizes the life-time of a route. In MDR the life-time of a path is dictated by the life-time of the bottleneck node along the path. The life-time of a node is calculated using the current Average-Energy-Drain-Rate (AEDR). Each node calculates the Energy-Drain-Rate (EDR), by calculating the energy consumed by the node for the last $T$ seconds ($T=6 \ seconds$ as specified by MDR authors in Section 3 of~\cite{mdr}). This computed EDR is averaged using exponential weighted moving average (with $\alpha=0.3$ as specified by MDR authors in Section 3 of~\cite{mdr}).

Let $DR_{i}{(t)}$ be the drain rate of node $i$ at time $t$ and $E_{r}^{i}{(t)}$ is the residual battery power of node $i$ at time $t$. Thus, the life-time of a path is determined by the minimum $T_{r}^{i}{(t)}$ along that path, where,
\begin{equation*}
T_{r}^{i}{(t)} = \frac{E_{r}^{i}{(t)}}{DR_{i}{(t)}}. 
\end{equation*}
Thus, the Minimum Drain Rate (MDR) mechanism selects the route with maximum life-time.

We have used AODV~\cite{aodv} as the underlying routing protocol and made necessary modifications for our MDR implementation. In~\cite{mdr}, authors have used Dynamic Source Routing (DSR) as underlying routing protocol for implementation of MDR, however authors claim that underlying routing protocol does not make a difference in the performance of the MDR scheme. Since both DSR and AODV are on-demand routing protocols and we have already used AODV for implemention of SQ-AODV, we decided to use AODV for MDR implementation.

For MDR routing protocol implementation, we have modified the RREQ message of AODV to carry the bottleneck life-time information of the path. As the RREQ message travels from source to destination, the bottleneck life-time field of the RREQ messages is updated. The destination node waits either for \textbf{$0.25 \ seconds$} after the first RREQ receiption or for the receiption of 3 RREQs, and finally reply to the RREQ that maximizes the life-time of the path. MDR updates its routes every $10 \ seconds$ to maintain up-to-date routing information. Hence a source node keep updating the routes by periodic route discoveries. We have simulated the AODV-based MDR, and present the simulation setup and the results in the Section~\ref{setup-mdr} and Section~\ref{results-mdr} respectively.

\subsection{Simulation Setup}
\label{setup-mdr}
\begin{figure}[htbp]
	\centering
	\includegraphics[scale=0.5]{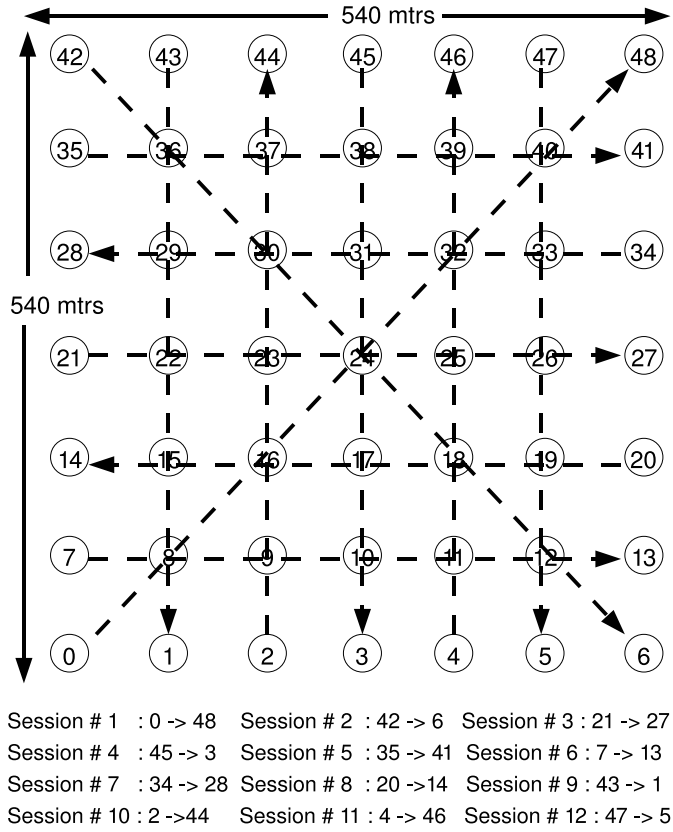}
	\caption{49 Node Topology with 12 Sessions}
	\label{fig57}
\end{figure}

We consider the 49-node static topology (where there is no node mobility) with 12 sessions as shown in Fig.~\ref{fig57} for our simulations. This is the same dense network scenario considered in~\cite{mdr}. The nodes are distributed uniformly in an area of size 540 m x 540 m, and are identical in their capability, but are initialized with different energies. The source and destinations have higher initial energy than other nodes, this is to make sure that the communication between source and destination starts.

The MAC layer protocol is IEEE 802.11 DCF (Distributed Co-ordination Function) with the PLCP (Physical Layer Convergence Protocol)  data rate being 1 Mbps. The parameters used for these simulations are listed in Table~\ref{tab54}. The data traffic in all the sessions is CBR, with data packets at each node arriving at 3 packets/sec and all the sessions starting at the start of the simulation. Initial energy of the source/destinations are chosen randomly between 50 to 75 Joules and that of intermediate nodes are selected between 25 to 50 Joules. The simulation was run 50 times (each initialized with a different seed), and the resulting parameters are averaged over these 50 runs.

\begin{table}[htbp]
\centering
  \caption{Values of Parameters Used for MDR Verification}
  \hfill \\	
  \begin{tabular}{|c|c|}
  \hline
  Packet size & 512 Bytes \\
  \hline
  Simulation time & 800 sec \\
  \hline
  Date traffic & CBR with 3 pkts/sec\\
  \hline
  MAC Protocol & IEEE 802.11 DCF \\
  \hline
  PCLP Data rate & 1 Mbps \\
  \hline
  Buffer length & 50 Packets \\
  \hline
  $P_{t}consume$ & 0.2818 W \\
  \hline
  $P_{r}consume$ & 0.2818 W \\
  \hline
  Propagation model & Two-Ray Ground \\
  \hline
  \end{tabular}
  \label{tab54}
\end{table}

\subsection{Results and Discussions}
\label{results-mdr}
\begin{figure}[htbp]
	\centering
	\includegraphics[scale=0.75]{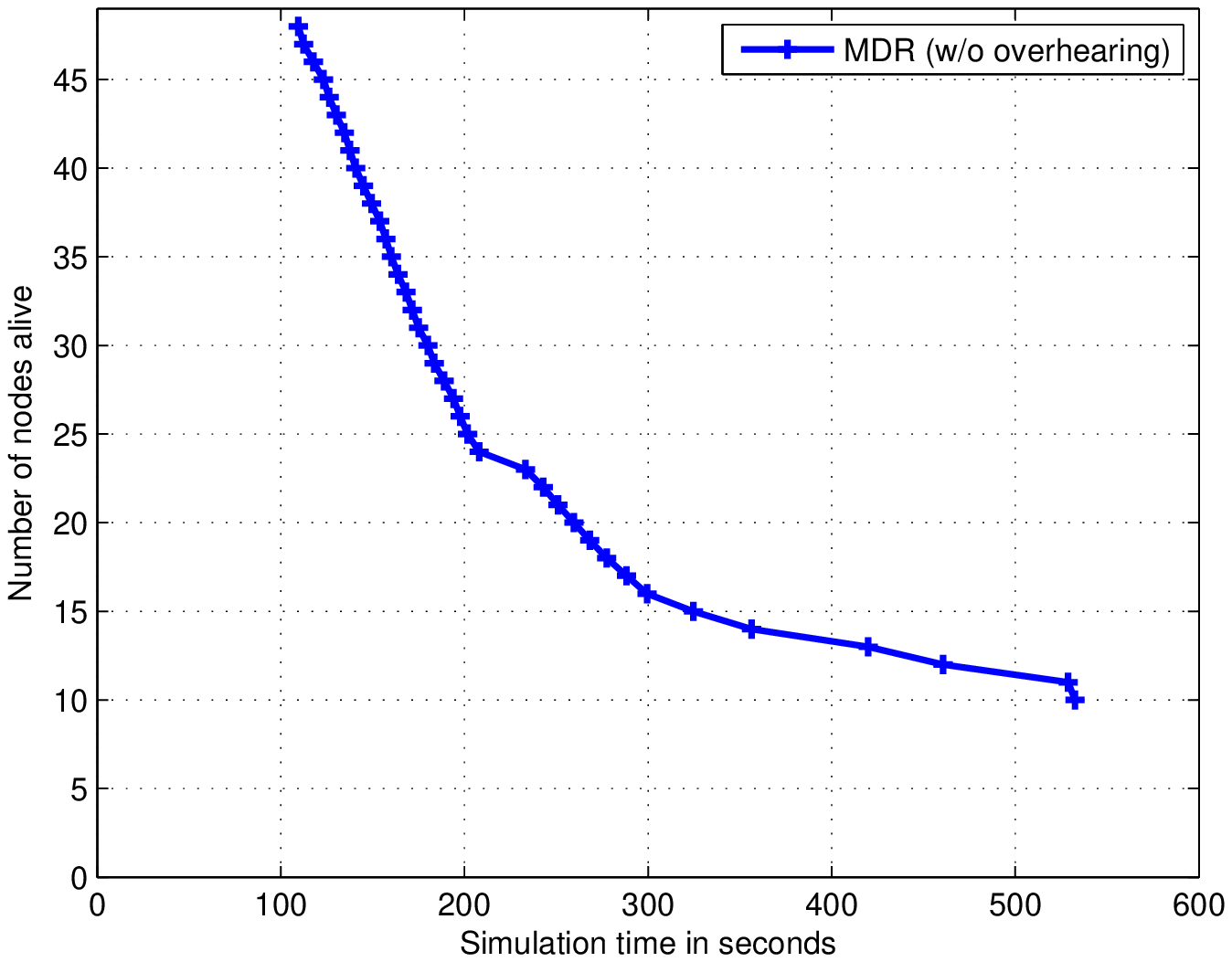}
	\caption{Node Expiration Time}
	\label{fig58}
\end{figure}

\begin{figure}[htbp]
	\centering
	\includegraphics[scale=0.75]{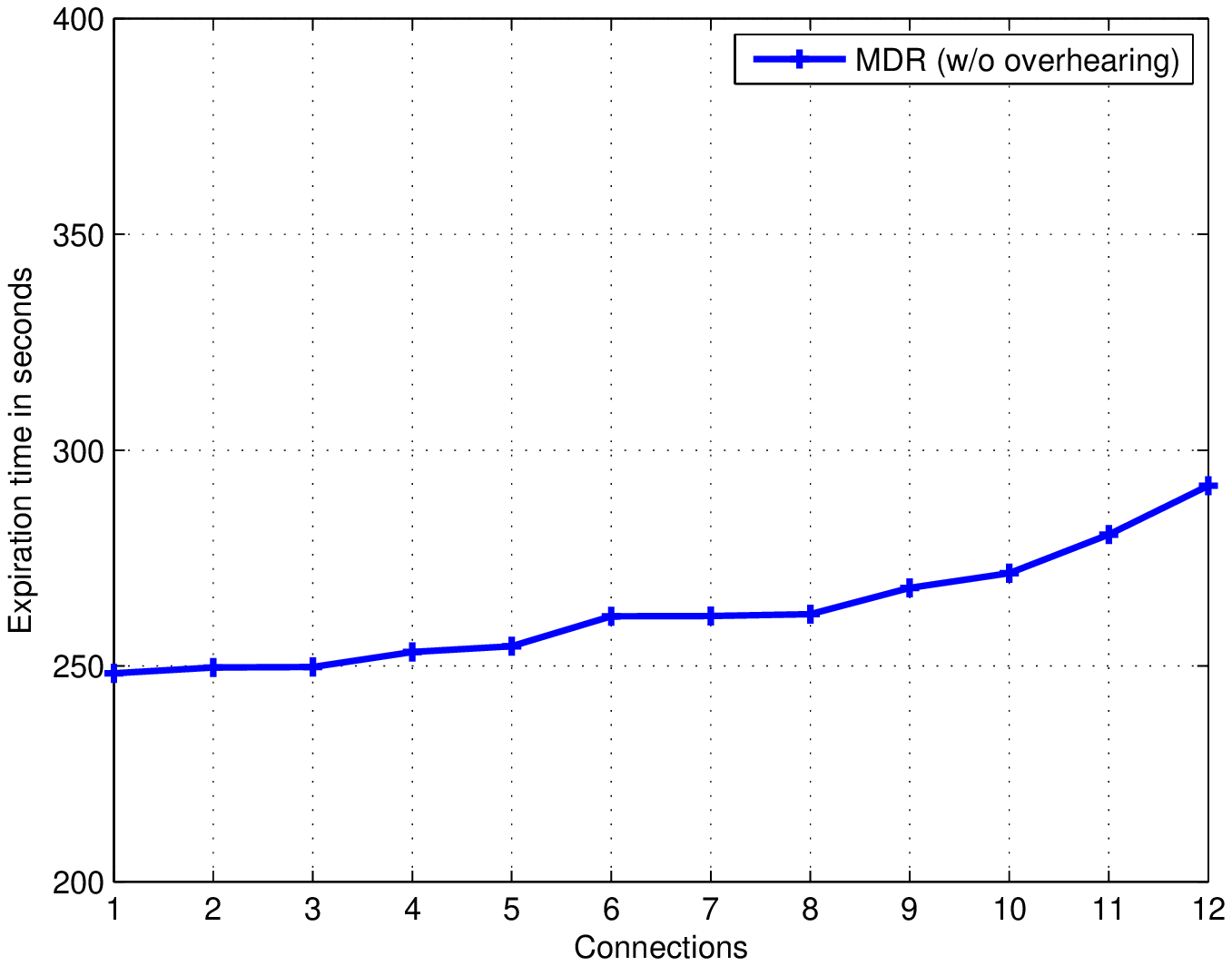}
	\caption{Connection Expiration Time}
	\label{fig59}
\end{figure}

\begin{figure}[htbp]
	\centering
	\includegraphics[scale=0.2]{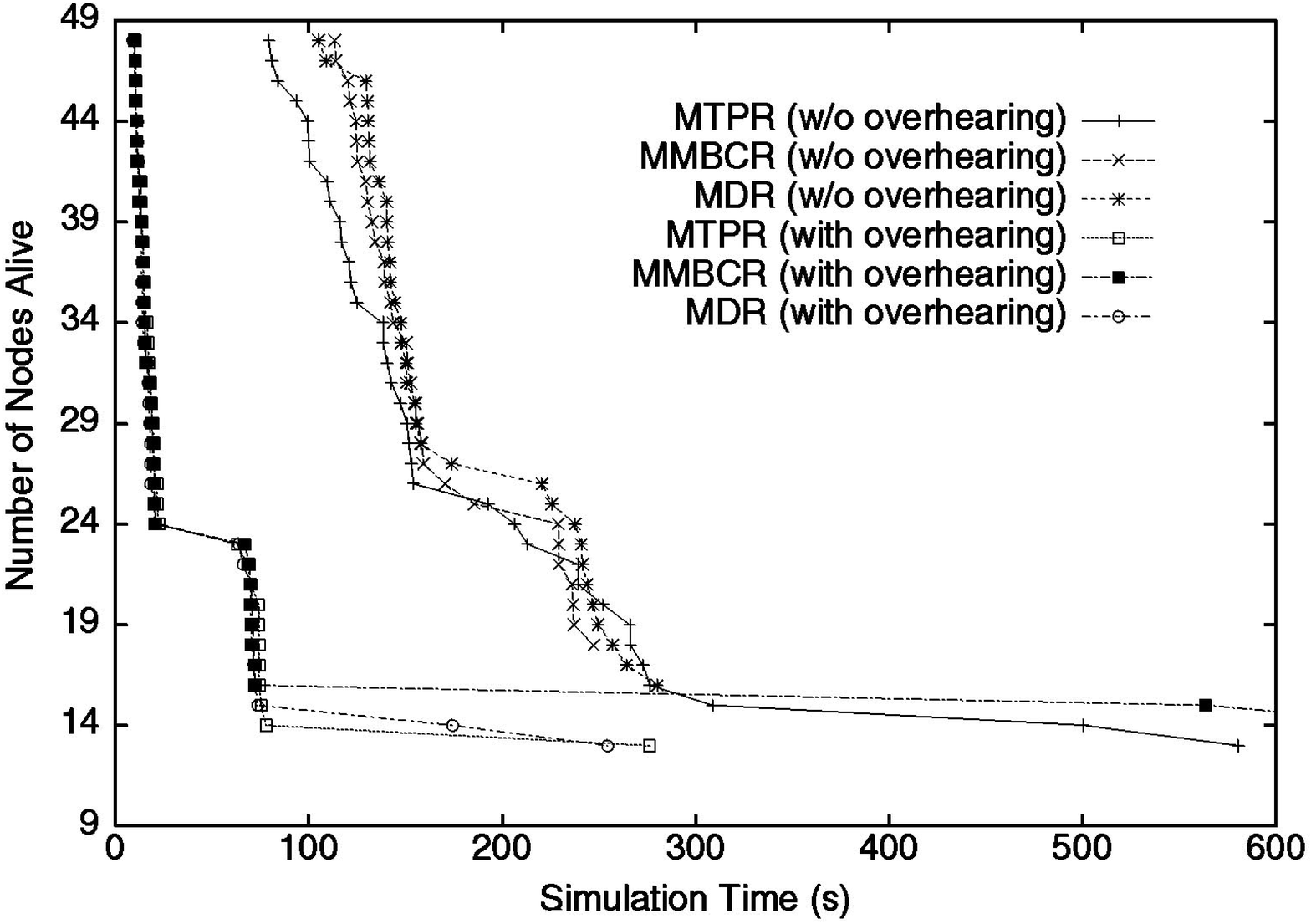}
	\caption{Node Expiration Time~\cite{mdr}}
	\label{fig510}
\end{figure}

\begin{figure}[htbp]
	\centering
	\includegraphics[scale=0.2]{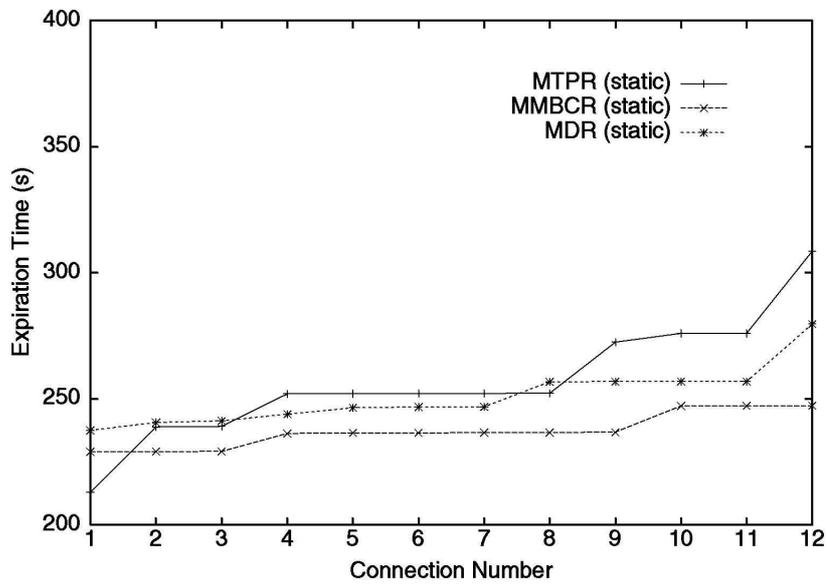}
	\caption{Connection Expiration Time~\cite{mdr}}
	\label{fig511}
\end{figure}

The results of Node Expiration Time and Connection Expiration Time, generated with our implementation of MDR routing protocol are plotted in Figure~\ref{fig58} and Figure~\ref{fig59} respectively and Figure~\ref{fig510} and Figure~\ref{fig511} are the results from the MDR paper~\cite{mdr} for Node Expiration Time and Connection Expiration Time respectively with * sign as sample points. Although authors of MDR paper~\cite{mdr} have not given enough information to duplicate their results, we have reverse engineered the network parameters to get close to their results. We have simulated an AODV-based MDR and the plots in Figure~\ref{fig58} and Figure~\ref{fig59} are qualitatively similar to those in origional MDR paper, this verifies the behavior of our implementation of MDR is similar to that of original MDR paper.

\section{Performance Comparison of SQ-AODV, MDR and AODV}
\label{comp-sq-aodv-mdr-and-aodv}
In this section, we present the simulation results to compare the performance of the three protocols SQ-AODV, MDR and AODV. For performance comparison, we have conducted two set of experiments, \textbf{Set A} and \textbf{Set B}. Set A is designed to evaluate the overall performance of SQ-AODV, MDR and AODV, for CBR traffic sources, while Set B is designed to evaluate the overall performance SQ-AODV, MDR and AODV, for Poisson traffic sources and at {\it varying network loads.} We present the results of these 2 sets along with the simulation setup in the next section.

\subsection{Simulation Setup for Set A}

We consider the 49-node static topology (where there is no node mobility) with 12 sessions as shown in Figure~\ref{fig57}, This is the same dense network scenario considered in~\cite{mdr}. The nodes are distributed uniformly in an area of size 540 m x 540 m, and are identical in their capability, but 
are initialized with different energies.

\begin{table}[htbp]
 \centering
  \caption{Values of Parameters Used for Simulation Set A} 
  \hfill \\
  \begin{tabular}{|c|c|}
  \hline
  Packet size & 512 Bytes\\
  \hline
  Simulation time & 800 seconds\\
  \hline
  Date traffic & CBR with 3 pkts/sec\\
  \hline
  MAC Protocol & IEEE 802.11 DCF\\
  \hline
  PCLP Data rate & 1 Mbps\\
  \hline
  Buffer length & 50 Packets\\
  \hline
  Transmit power & 0.2818 W\\
  \hline
  Propagation model & Two-Ray Ground\\
  \hline
  \end{tabular}
  \label{tab55}
\end{table}

Simulation Set A involves 2 experiments. In the first, all sessions begin transmission at the start of the simulation, and the simulation runs for a fixed duration (800 s). In the second, the session start times are chosen 
randomly. In both experiments, the initial energy of the nodes is uniformly distributed between 25 J and 100 J, and data at each node arrives at
3 pkts/sec or about 12 Kbps. Every experiment was run 50 times (each initialized with a different seed), and the resulting parameters averaged over these 50 runs. The network parameters used in simulation Set A are detailed in Table~\ref{tab55}.

\subsection{Results and Discussions for Set A}
\label{results-seta}

\begin{figure}[htbp]
	\centering
	\includegraphics[scale=0.75]{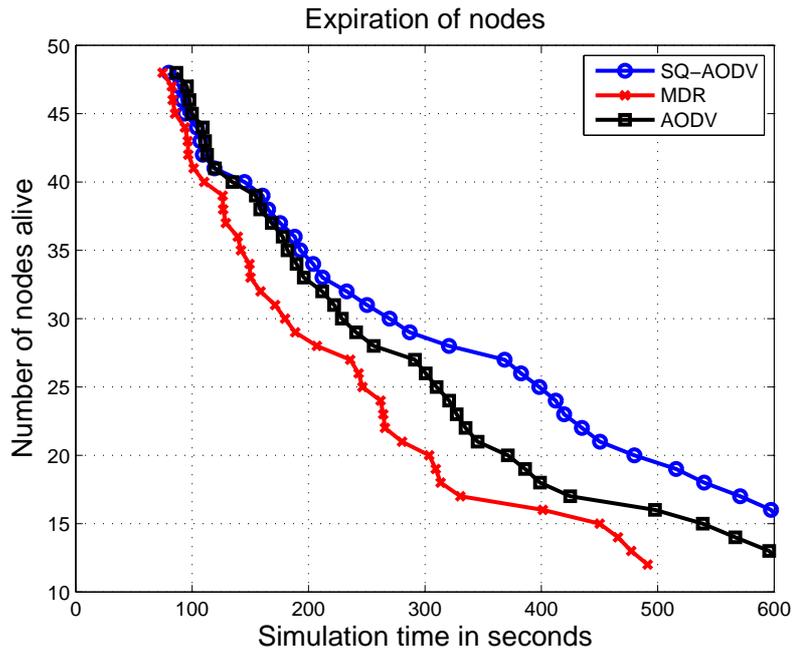}
	\caption{NET: Sessions Commence at Start of Simulation}
	\label{fig512}
\end{figure}

\begin{figure}[htbp]
	\centering
	\includegraphics[scale=0.75]{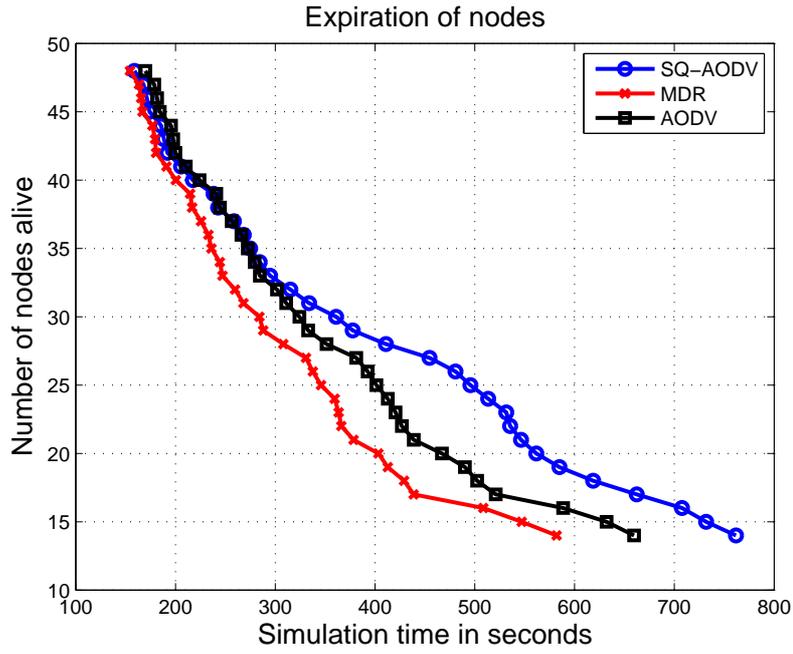}
	\caption{NET: Random Session Start Times}
	\label{fig513}
\end{figure}

\begin{figure}[htbp]
	\centering
	\includegraphics[scale=0.75]{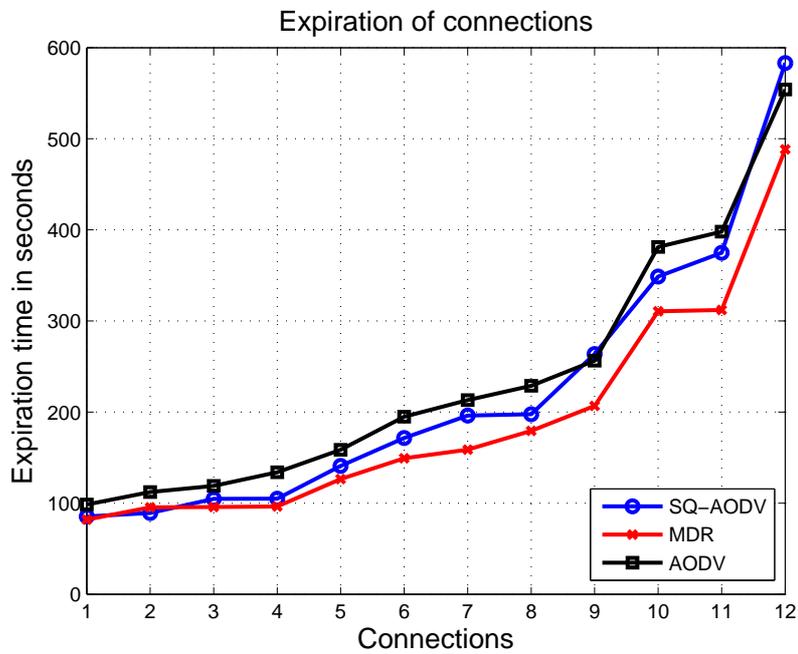}
	\caption{CET: Sessions Commence at Start of Simulation}
	\label{fig514}
\end{figure}

\begin{figure}[htbp]
	\centering
	\includegraphics[scale=0.75]{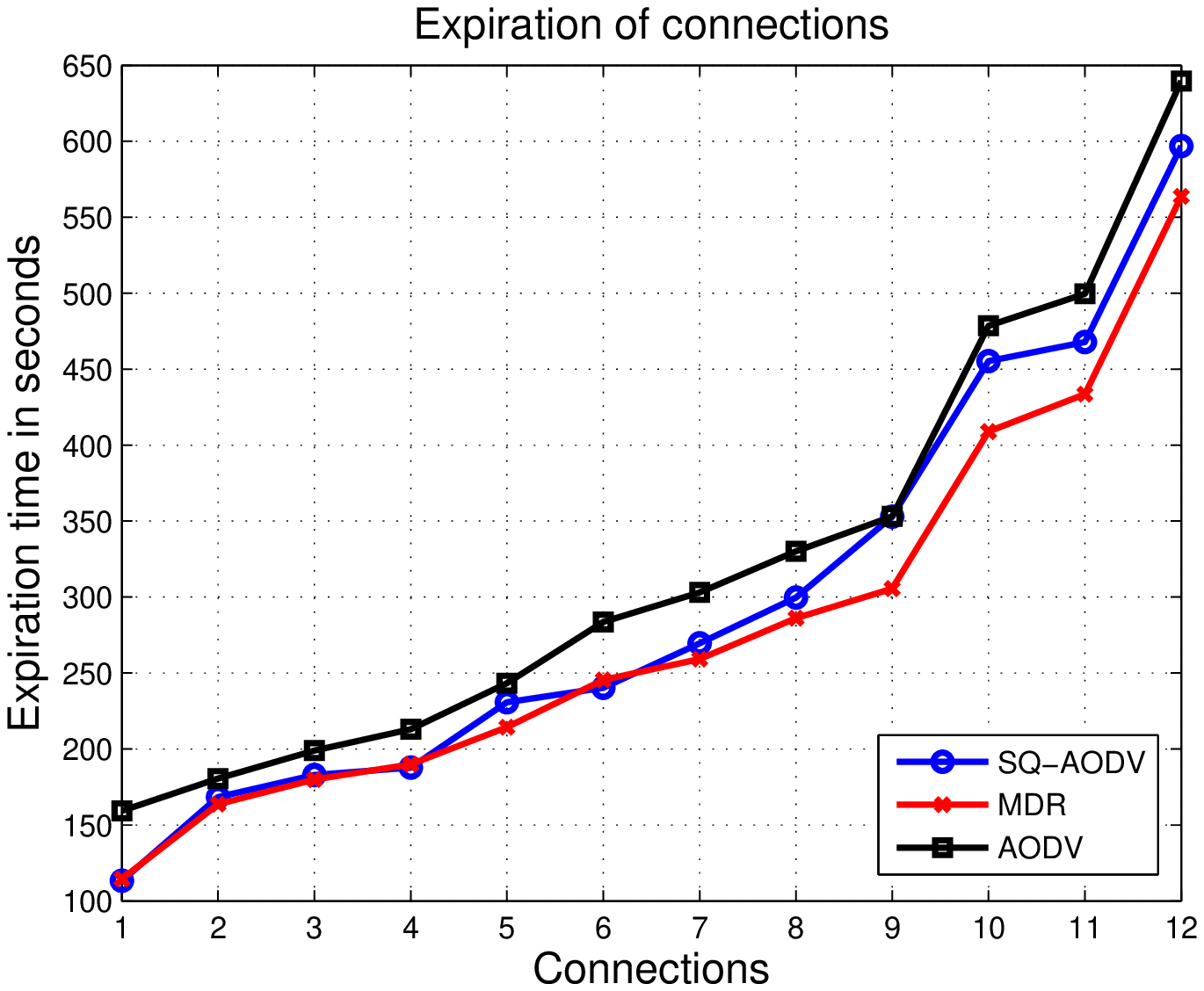}
	\caption{CET: Random Session Start Times}
	\label{fig515}
\end{figure}

The results of the two experiments from simulation Set A are presented in Table~\ref{tab56}, while the plots of NET and CET are presented in Figs.~\ref{fig512} -~\ref{fig515}

\begin{table}[htbp]
\centering
\caption{Simulation Results for Set A}
\hfill \\
\begin{tabular}{|c|c|c|c|c|c|c|}
\hline
Parameter  & \multicolumn{3}{c|}{Set-A(1)} & \multicolumn{3}{c|}{Set-A(2)} \\
\        & SQ-AODV & MDR & AODV & SQ-AODV & MDR & AODV  \\
\hline
PDR            & 0.9760   & 0.8456   & 0.8681 & 0.9892   & 0.9201   & 0.8926 \\
\hline
COH            & 0.7742   & 13.3207  & 1.0877 & 0.3402   & 4.1554   & 0.8256 \\
\hline
PD (sec)       & 0.0618   & 0.2429   & 0.0543 & 0.0348   & 0.0508   & 0.0353 \\
\hline
\end{tabular}
\label{tab56}
\end{table}

We see from Table~\ref{tab56} that the PDR for SQ-AODV in the two experiments is improved by 12.5\% and 10.8\%, respectively, relative to AODV. This is because choosing nodes with limited battery life, as happens in AODV, leads to (avoidable) disconnections of sessions. SQ-AODV, on the other hand, performs better because: (i) it chooses those intermediate nodes whose energy is sufficient to support the session for its entire duration or it chooses nodes to maximize the life-time of the route, and (ii) due to its make-before-break strategy, which re-routes a session proactively when link failure due to depletion of node energy is imminent. Thus, SQ-AODV successfully reduces packet drops in the network quite significantly.

Similarly, the PDR in MDR in the two cases is poorer by 15.4\% and 7.5\%, respectively as compared to SQ-AODV. This is because MDR's periodic route update feature adds substantial routing overhead in the network. In fact,  Table~\ref{tab56} shows that MDR overhead is approximately 17 times and 12 times worse than that of SQ-AODV, respectively, and  almost 12 times and 5 times worse than that of AODV, respectively. This leads to its much lower PDR.

The packet delay for both SQ-AODV and AODV is comparable in both cases. We posit that this is because the delay in SQ-AODV is the result of two opposing factors. On the one hand, finding stable routes, where the life-time of the bottleneck node is maximized, may lead to longer (but more stable) routes,
thus increasing delay. On the other hand, proactive route maintenance by way of make-before-break decreases delay, since no retransmissions need occur while an alternative route is located. These two factors have a compensatory effect, making the packet delay in SQ-AODV of the same order as that in AODV. Results in Section~\ref{results-setb} (Table~\ref{tab58}) demonstrate the compensatory effects which makes the delay in SQ-AODV and AODV comparable. MDR, by contrast, imposes a much higher load on the network due to its periodic route updates making the data packets to wait longer, leading to a delay that is about 4 times and 1.5 times, respectively, the delay for AODV or SQ-AODV. 

We see from Figs.~\ref{fig512} and~\ref{fig513} that in our network setting, running SQ-AODV improves the node expiration time by between 25 to 100 seconds over AODV, and by between 100 to 150 seconds over MDR. In other words, for a given number of nodes alive, this equates to SQ-AODV extending the node life-time by between 10\%-25\% over AODV, and by between 25\%-35\% over MDR. Viewed another way, at a given simulation time, SQ-AODV typically has between 10\%-25\% more nodes alive than does AODV, and has between 20\%-60\% more
nodes alive than does MDR. This is because SQ-AODV's proactive route maintenance is very economical of node energy. In addition, due to the proactive mechanism in SQ-AODV, a source stops transmitting traffic if a destination is about to drain, which saves resources by minimizing the transmission of packets that would not have been received by the destination in any case (due to its expiring). The nodes in MDR, on the other hand, expire 
faster than they do in either AODV or SQ-AODV by a significant margin, this is because the periodic updates of MDR consume energy at a substantially
faster rate, causing nodes to expire much quicker, as our results demonstrate.

Figs.~\ref{fig514} and~\ref{fig515} illustrate that in our network setting, in terms of CET, AODV performs better by about 10-50 seconds over both SQ-AODV and MDR. (Note that the x-axis in these figures simply indicates the {\it number of connections} that have expired, and is {\it not} the connection identifier . Thus, the connections that expire under each protocol can be different. So, for example, Fig.~\ref{fig514} shows that 6 connections expire at approximately 200 seconds with AODV, at 180 seconds with SQ-AODV, and at 165 seconds with MDR. However, the 6 sessions that expire under each protocol
are not the same sessions).

This equates to AODV connection expiration times being anywhere between 30\%-7\% better than those of SQ-AODV or MDR. This is because, in SQ-AODV: (i) a source on receiving an RCR from an intermediate node tries only a fixed (but configurable; in our case 3) times before it reaches the maximum number of
retries and ends the session, and (ii) intermediate nodes reject a new session once its residual-energy is bellow Threshold-1. By contrast AODV keeps retrying and so has a higher probability of finding a path, and keeping the session alive for longer. In the case of MDR, however, it is the control overhead packets that cause the node energy to drain faster, leading to the sessions expiring quicker than with AODV or SQ-AODV. We note that the slightly higher connection expiration times in AODV do come at the cost of lower PDR and lower node expiration times, which implies that even though the connections may be alive for a longer period in AODV, they do not successfully transmit as much data as SQ-AODV does.

\subsection{Simulation Setup for Set B}

We consider the same 49-node static topology (where there is no node mobility) with 12 sessions as shown in Figure~\ref{fig57} for Set B. The nodes are distributed uniformly in an area of size 540 m x 540 m, and are identical in their capability, but are initialized with different energies. In this set The traffic arrives as per a Poisson process, for different network loads. The initial node energies are uniformly distributed between 75 J and 300 J, and the simulation is run until each session has transmitted 3000 packets, while the session data rates vary from 15 Kbps to 65 Kbps. Here we again examine performance by comparing PDR, control overhead (COH) and packet delay (PD) at {\it{different network loads}}.

\begin{table}[htbp]
 \centering
  \caption{Parameters and Their Values Used for Simulation Set B} 
  \hfill \\
  \begin{tabular}{|c|c|}
  \hline
  Packet size & 512 Bytes\\
  \hline
  Packets/Session & 3000 \\
  \hline
  Date traffic & Poisson with $\lambda$ = 15 Kbps - 65Kbps\\
  \hline
  MAC Protocol & IEEE 802.11 DCF\\
  \hline
  PCLP Data rate & 1 Mbps \\
  \hline
  Buffer length & 50 Packets \\
  \hline
  Transmit power & 0.2818 W\\
  \hline
  Propagation model & Two-Ray Ground\\
  \hline
  \end{tabular}
  \label{tab57}
\end{table}

Every experiment was run 50 times (each initialized with a different seed), and the resulting parameters averaged over these 50 runs. The network parameters used in simulation Sets A and B are detailed in Table~\ref{tab57}.

\subsection{Results and Discussions for Set B}
\label{results-setb}
\begin{figure}[htbp]
	\centering
	\includegraphics[scale=0.75]{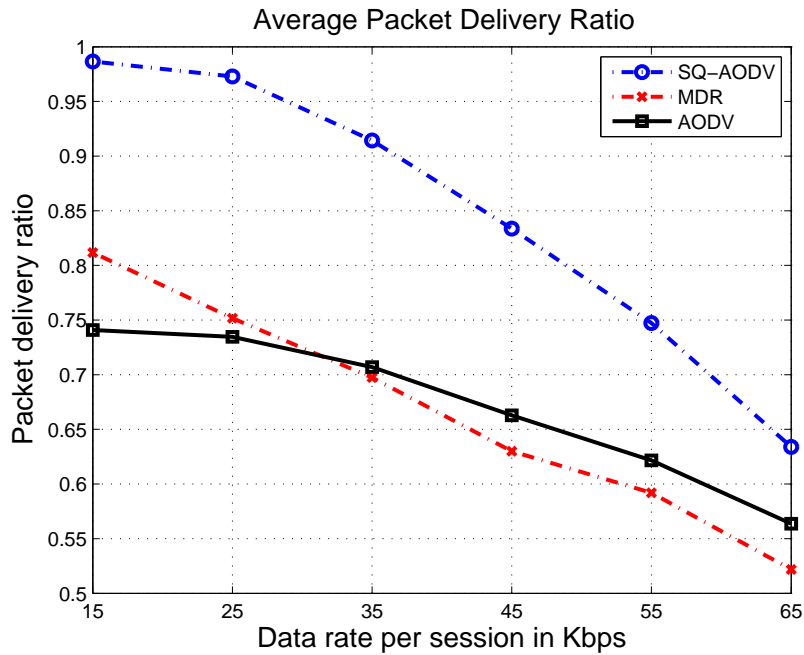}
	\caption{Average Packet Delevery Ratio for Simulation Set B}
	\label{fig516}
\end{figure}

We observe from Fig.~\ref{fig516}, the PDR of SQ-AODV is substantially better than that for AODV or MDR. In fact, the PDR for SQ-AODV is improved by between 25\%-13\% over AODV and by between 22\%-18\% over MDR over the network loads considered. The key reason for this are  the two properties of SQ-AODV discussed in Chapter~\ref{intro-sq-aodv}, which induce stable routes for the sessions and bolster PDR.

The PDR of MDR is better than that of AODV by 5\%-10\% at lower loads, but is reduced by equal amount 
at higher loads because of its extra overhead, which degrades MDR performance at higher network loads.

\begin{figure}[htbp]
	\centering
	\includegraphics[scale=0.75]{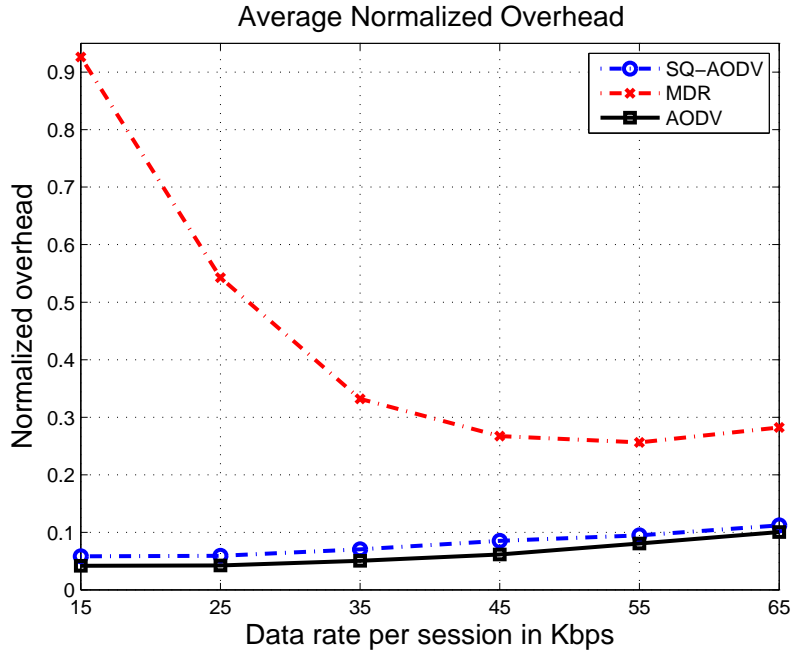}
	\caption{Average Control Overhead for Simulation Set B}
	\label{fig517}
\end{figure}

Fig.~\ref{fig517} shows that SQ-AODV has marginally higher normalized control overhead (between 1\%-3\% higher) than AODV. This is because, as explained in Chapter~\ref{intro-sq-aodv}, to support stable routing, SQ-AODV uses per-session (or per-flow) based routing (as opposed to simple destination-based routing used in AODV). For this, control packets of SQ-AODV carry extra flow-id information along with source and destination, and also packets need travel all the way to destination to find a stable path, leading to a marginally higher control overhead.

We see that MDR has the highest control overhead, almost 300\% higher than either AODV or SQ-AODV at loads above 35 Kbps, rising to over 1000\% higher at lower loads. This is because of the control overhead of MDR. In particular,
at lower loads the control overhead of MDR becomes very high, because it takes substantial time for the sources to generate 3000 packets. In the meantime, the regular  periodic update procedures of MDR continue accumulating significant control overhead.

\begin{figure}[htbp]
	\centering
	\includegraphics[scale=0.75]{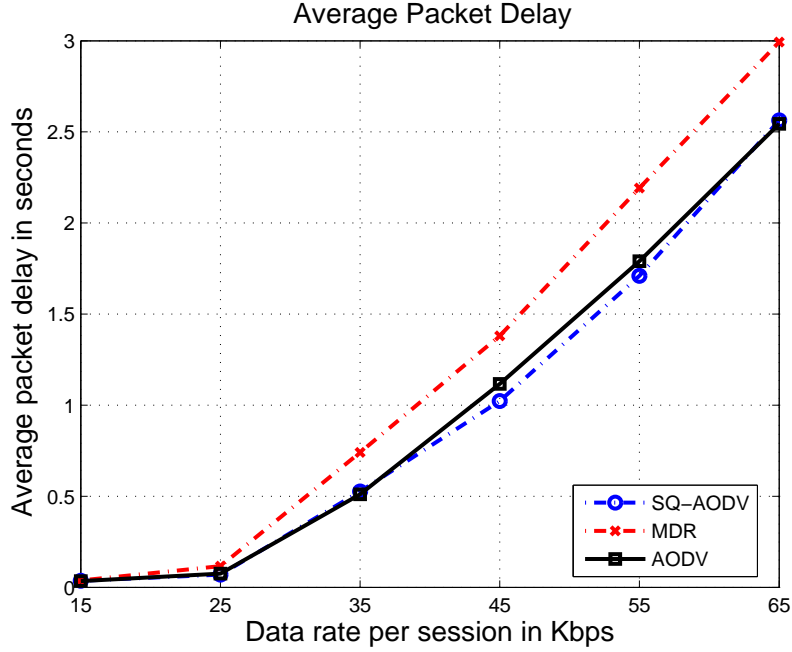}
	\caption{Average Packet Delay for Simulation Set B}
	\label{fig518}
\end{figure}

Finally Fig.~\ref{fig518} illustrates that, the delay experienced by packets in SQ-AODV is almost the same or marginally better than that in AODV, at all loads under consideration. This is because the delay in SQ-AODV is the result of two opposing factors. On the one hand, finding stable routes, where the life-time of the bottleneck node is maximized, may lead to longer (but more stable) routes, thus increasing delay. On the other hand, proactive route maintenance by way of make-before-break decreases delay, since no retransmissions need occur while an alternative route is located. These two factors have a compensatory effect, making the packet delay in SQ-AODV of the same order as that in AODV. The delay experienced in MDR is between 250-500 ms higher, or between 20\%-50\% higher than that in AODV and SQ-AODV because, at higher loads data packets have to wait longer due to periodic route updates.
The advantage with SQ-AODV is that it is designed to provide stable routes and a fast re-routing capability to the nodes in ad-hoc networks at minimum overhead to the network. This helps in making effective use of network resources, as demonstated by our simulation results.

\begin{table}[htbp]
\centering
\caption{Average Number of Hops for Simulation Set B}
\hfill \\
\begin{tabular}{|c|c|c|c|c|c|c|}
\hline
Protocol & \multicolumn{6}{c|}{Session Data Rate}\\
\	 & 15 Kbps & 25 Kbps & 35 Kbps & 45 Kbps & 55 Kbps & 65 Kbps\\
\hline
SQ-AODV  & 4.9503 & 5.0165 & 5.0519 & 5.0487 & 5.0938 & 5.2244 \\
\hline
MDR      & 5.0147 & 5.0258 & 5.0800 & 5.1616 & 5.0861 & 5.1326 \\
\hline
AODV     & 4.5294 & 4.4972 & 4.5858 & 4.6581 & 4.6984 & 4.7311 \\
\hline
\end{tabular}
\label{tab58}
\end{table}

Table~\ref{tab58} gives the number of hops taken by the data packets to travel from source to destination in SQ-AODV, MDR and AODV, on an average. As we can see the number of hops taken by SQ-AODV and MDR is increased by about 11\% over AODV on an average over all loads. This should clearly indicate that the packet delay in case of SQ-AODV would have been more, but the proactive route maintenance by way of make-before-break decreases delay in SQ-AODV, leading to an almost similar delay in both SQ-AODV and AODV over all loads. On the other hand for MDR, the data packets experience a larger delay compared to both SQ-AODV and AODV as explained earlier in Section~\ref{results-seta}.

%% file: conclusion.tex
\chapter{Summary and Future Work}
\label{conclusion}
In this thesis, we have proposed a novel, energy-aware, stable routing protocol named, Stability-based QoS-capable Ad-hoc On-demand Distance Vector (SQ-AODV) protocol which is an enhancement of the well-known AODV protocol for ad-hoc wireless networks. SQ-AODV utilizes a cross-layer design approach in which information about the residual energy of a node is used for route selection and maintenance. An important feature of SQ-AODV is that it uses only local information and requires no additional communication or co-operation between the network nodes. SQ-AODV has a proactive route maintenance by make-before-break which increases the packet delivery ratio in the network at virtually no extra overhead, making it more suitable for ad-hoc wireless environment. SQ-AODV is also compatible with the basic AODV data formats and operation, making it easy to adopt in ad-hoc networks.

Simulation results shows under variety of applicable network loads and network parameters, SQ-AODV protocol acheives packet delivery ratio, on an average, 10-15\% better than either AODV or Minimum Drain Rate (MDR) routing protocol, and node expiration times 10-50\% better than either AODV or MDR, with packet delay and control overhead comparable to that of AODV.

Several directions of future work are possible from here. The first is to combine our scheme explicitly with QoS routing, thereby incorporating bandwidth and delay constraints in the path selection process. Another is to consider the effects of mobility and fading in our stable routing protocol.

%% file: app1.tex
\chapter{Introduction to NS-2}
\label{intro-ns-2}
In this appendix, we give a very brief introduction to NS-2. This introduction helps a reader to get a basic orientation to NS-2 and enable the reader to better understand the implementation details of SQ-AODV given in Appendix~\ref{sq-aodv-details-in-ns-2}.

NS-2 is an object-oriented, discrete event driven network simulator written in C++ and OTcl (Object-oriented Tool command language). NS-2 provides substantial support for simulating wired and wireless networks with network protocols such as TCP and UDP. NS-2 has support for, traffic source behavior such as FTP, Web, CBR and VBR, router queue management mechanism such as Drop Tail, RED and CBQ, standard routing protocols and Link layer protocols for both wired and wireless networks.

Although NS-2 is very easy to use once you get to know it, but is quite difficult for a beginner. To get a feel of what is NS-2, a beginner is recommended to exercise some of the examples given in~\cite{NS-BY-EX}. One can also find a more detailed documentation of NS-2 in~\cite{NS-DOC}, which is written with the depth of a skilled user.

\section{C++ and OTcl Linkage in NS-2}
\begin{figure}[htbp]
	\centering
	\includegraphics[scale = 0.45]{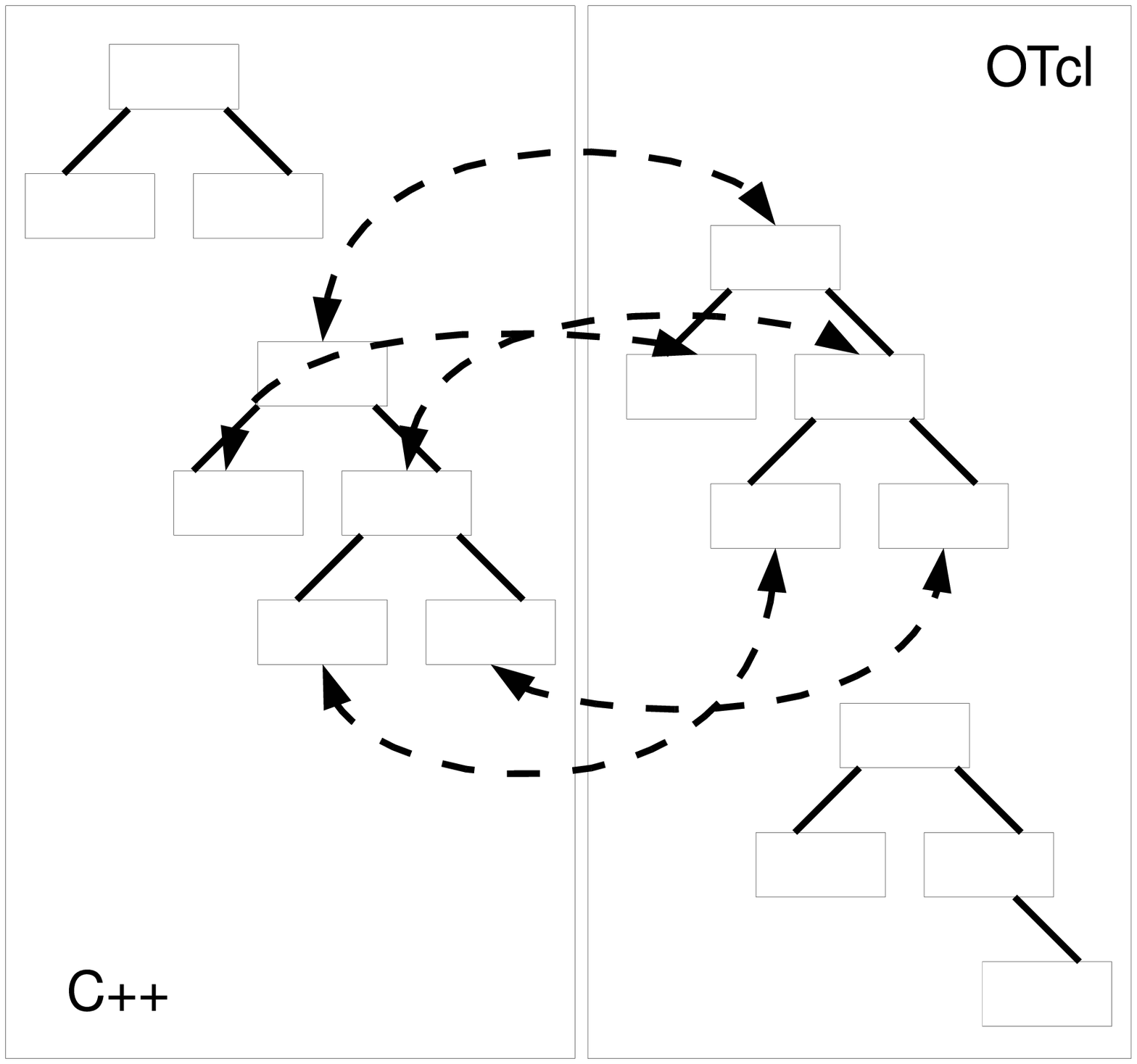}
	\caption{C++ and OTcl Linkage in NS-2}
	\label{fig41}
\end{figure}

NS-2 is written in both C++ and OTcl languages, with data path using C++ and control path using OTcl. In order to reduce the packet and event processing time, the event scheduler and the basic network component objects are written and compiled using C++. These compiled objects are made available to OTcl interpreter through OTcl linkage as shown in Fig.~\ref{fig41}.

The OTcl linkage creates a matching OTcl object for each of the C++ objects. Thus giving the control of the C++ objects to OTcl. There are objects in C++ that do not need OTcl control, these objects are not in OTcl space. Similarly there are objects that are entirely implemented in OTcl and are not in C++ space. Thus, there is a matching OTcl object hierarchy very similar to that of C++. We now give a very brief explanation as how NS works.

\section{Simplified User's View of NS-2}
\begin{figure}[htbp]
	\centering
	\includegraphics[scale = 0.45]{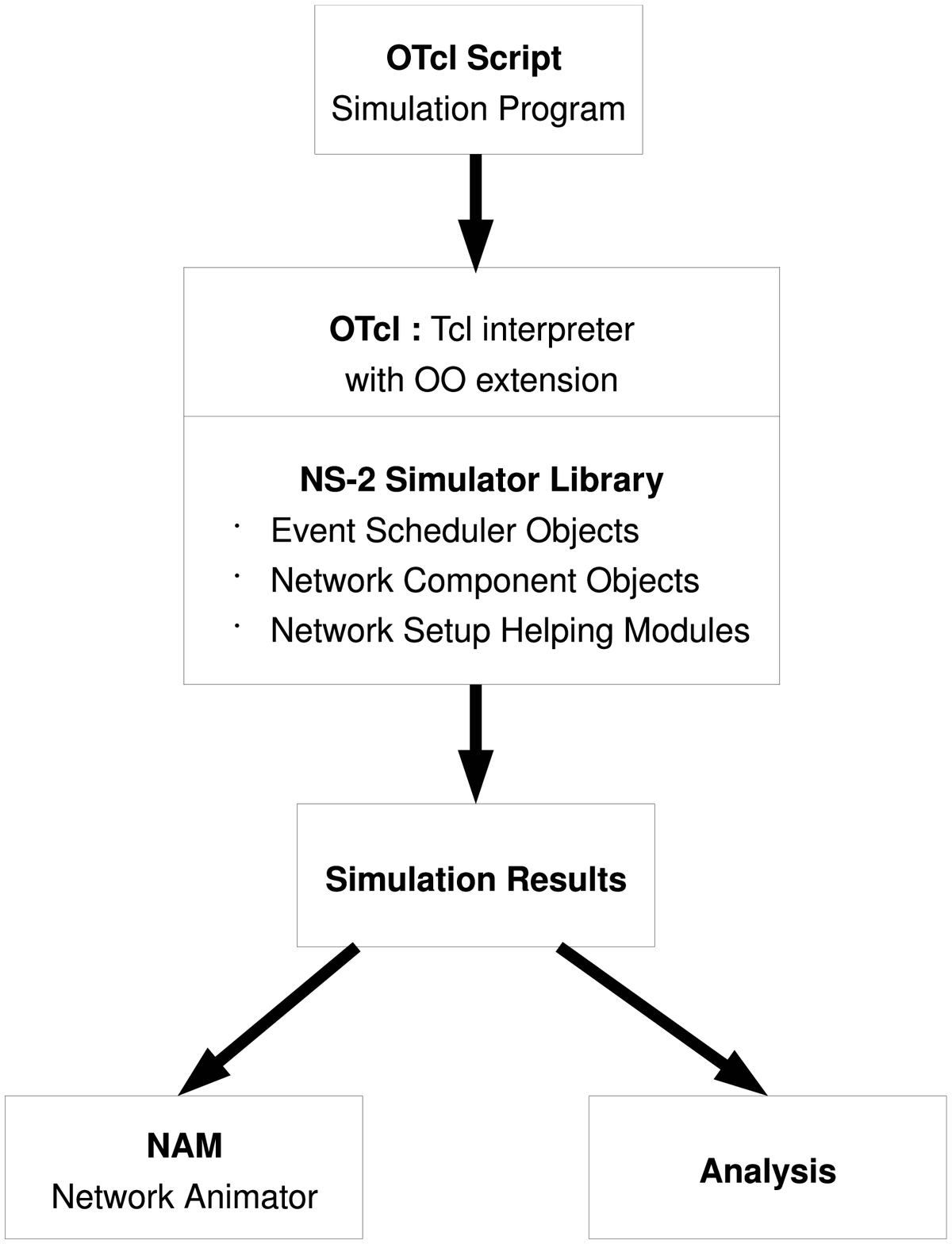}
	\caption{Simplified User's View of NS-2}
	\label{fig42}
\end{figure}

As shown in Fig.~\ref{fig42}, NS-2 is Object-oriented Tcl (OTcl) script interpreter that has a simulation event scheduler and network component object libraries, and network setup helping module libraries. The procedure for simulating a network scenario using NS-2 is as follows: User has to write a program using OTcl script language. This program basically initiates an event scheduler, sets up a network topology using network component objects and helping modules, and tell the traffic sources when to start and stop the transmission of packets through the event scheduler. This OTcl script is executed by NS-2 to generate the trace of events. The event scheduler will keep track of simulation time and dump all the events of the event queue scheduled at the current time, into an output file to generate the trace file output. Thus generated trace file is further processed by user written scripts to analyze the simulation results. Alternatively, the simulation results can also be visualized using Network AniMater (NAM)

%% file: app2.tex
\chapter{Details of SQ-AODV in NS-2}
\label{sq-aodv-details-in-ns-2}
In this appendix, we provide implementation details of SQ-AODV in Network Simulator Version-2~\cite{NS-2} (NS-2). We first give the changes made to the AODV routing protocol packet formats, and then give the changes made to the functions of AODV to convert them into corresponding functions of SQ-AODV.

\section{Format of Routing Packets in AODV and SQ-AODV}
\begin{figure}[htbp]
	\centering
	\includegraphics[scale = 0.5]{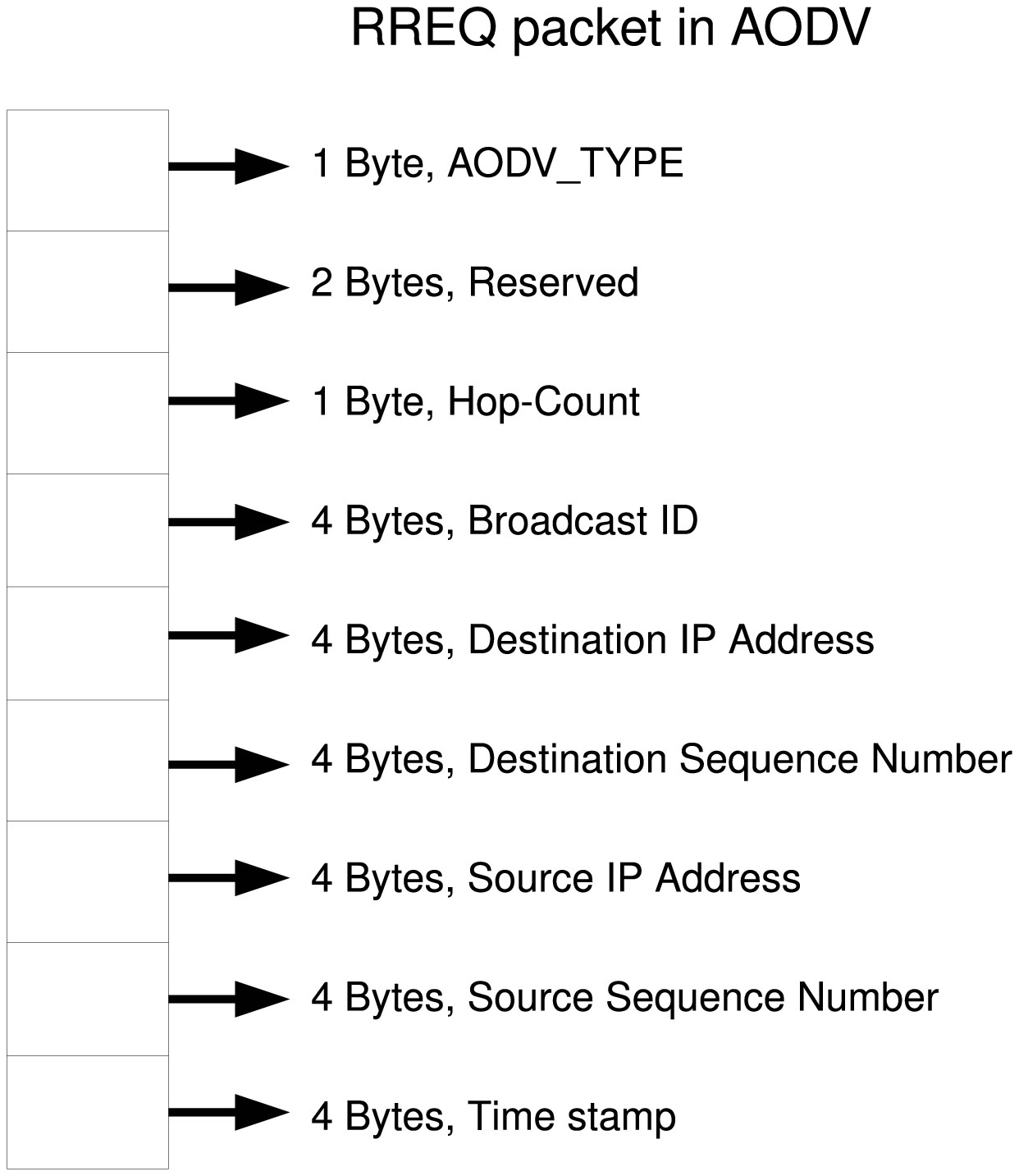}
	\caption{RREQ Packet Format in AODV }
	\label{fig421}
\end{figure}

The format of the RREQ is as shown in the Figure~\ref{fig421}, and contains the following fields:
\begin{itemize}
\item \textbf{Type} -- Indicates the type of the packet to differentiate between RREQ, RREP, ERROR packets

\item \textbf{Hop count} -- The number of hops from originator node to the node handling the request

\item \textbf{BroadcastID} -- A sequence number uniquely identifying the particular RREQ packet when taken in conjunction with originating node's IP address

\item \textbf{Destination IP address} -- IP address of the destination for which a route is desired

\item \textbf{Destination sequence number} -- The latest sequence number received in the past by the originator for any route towards the destination

\item \textbf{Source IP address} -- The IP address of the node which originated the RREQ

\item \textbf{Source sequence number} -- The current sequence number to be used in the route entry pointing towards the source of the route request
\end{itemize}

\begin{figure}[htbp]
	\centering
	\includegraphics[scale = 0.5]{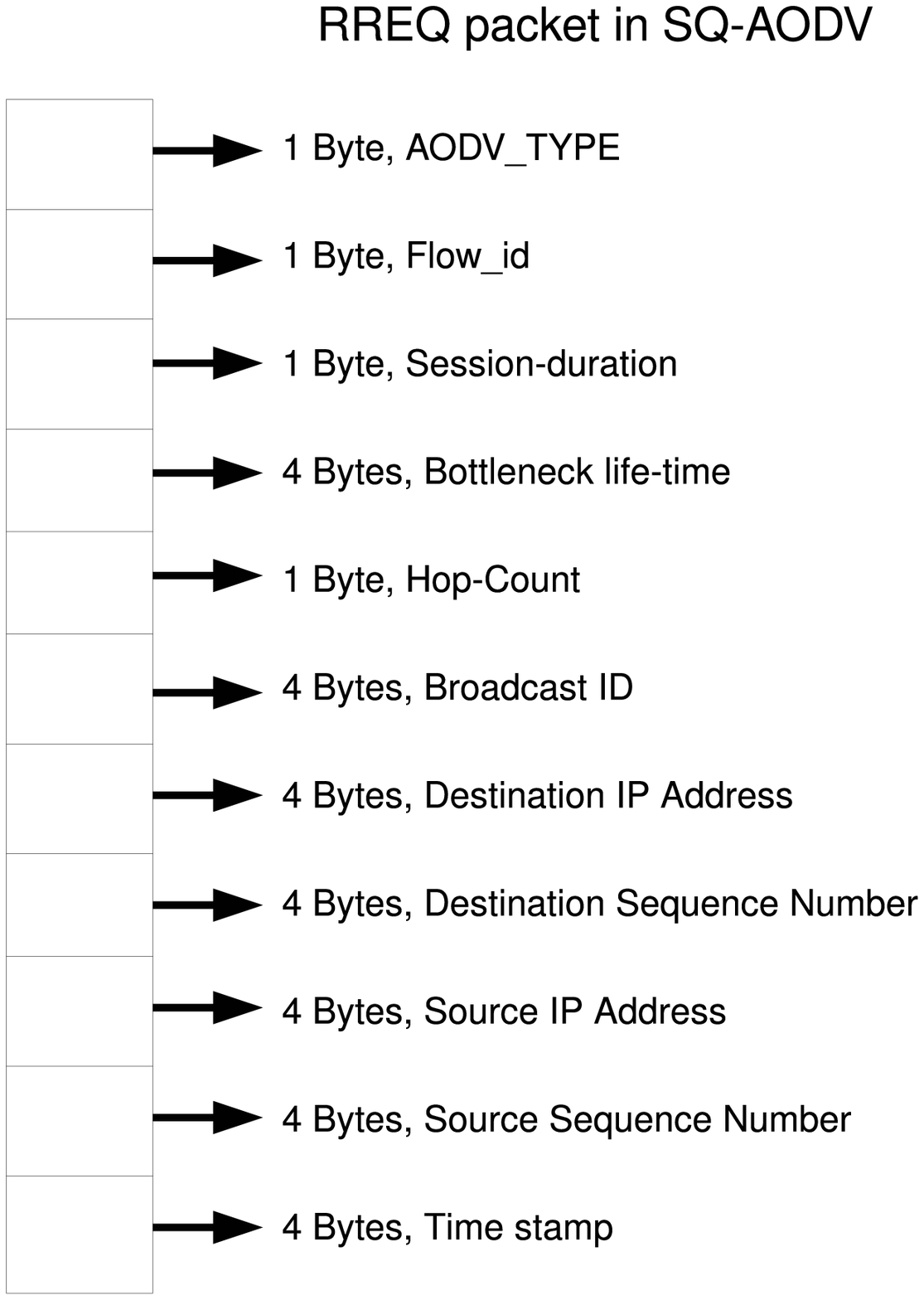}
	\caption{RREQ Packet Format in SQ-AODV }
	\label{fig422}
\end{figure}

The RREQ packet format modified for SQ-AODV is as shown in Figure~\ref{fig422} and the additional fields added are: 
\begin{itemize}
\item \textbf{Flow$\_$id} -- A sequence number uniquely identifying a flow (for which a route is desired) when taken in conjunction with the originator and the destinations IP address

\item \textbf{Session-duration} -- The session-duration information provided by the application of a flow

\item \textbf{Bottleneck life-time} -- The life-time of the bottleneck node along the path taken by the RREQ packet
\end{itemize}

In SQ-AODV routing is per session based as against destination-based in AODV. A route is uniquely identified by the triple $\{$Src, Dst and Flow$\_$id$\}$, hence the RREQ packet carries the flow$\_$id information to facilitate per session based routing. Session-duration is one of the parameters used to admit a new flow, hence this information is also need to be carried in RREQ packets to find a route that can survive atleast for the duration of the session. For flows not specifying the session-duration, SQ-AODV finds a stable path by selecting a path that maximizes the bottleneck node life-time along the path, to facilitate this a field is added to carry the bottleneck life-time of the node along a path.

\begin{figure}[htbp]
	\centering
	\includegraphics[scale = 0.5]{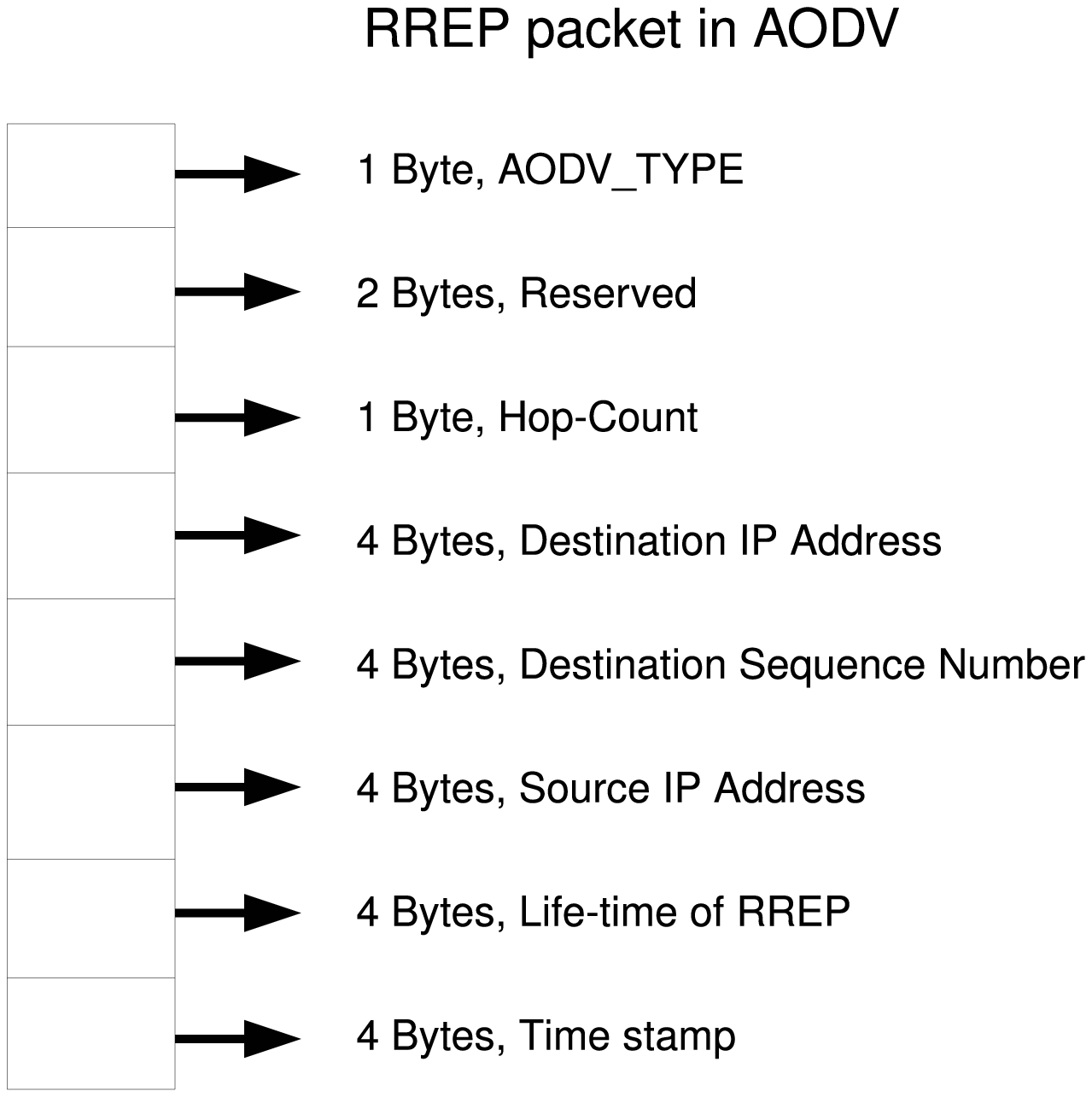}
	\caption{RREP Packet Format in AODV }
	\label{fig423}
\end{figure}

The format of the RREP is as shown in the Figure~\ref{fig423}, and contains the following fields:
\begin{itemize}
\item \textbf{Type} -- Indicates the type of the packet to differentiate between RREQ, RREP, ERROR packets

\item \textbf{Hop count} -- The number of hops from originator node to the node handling the request

\item \textbf{Destination IP address} -- IP address of the destination for which a route is desired

\item \textbf{Destination sequence number} -- The latest sequence number received in the past by the originator for any route towards the destination

\item \textbf{Source IP address} -- The IP address of the node which originated the RREQ

\item \textbf{Life-time} -- The time for which nodes receiving the RREP consider the route to be valid

\item \textbf{Timestamp} -- The time at which the RREP packet has been sent by a node (either destination or intermediate) towards the source requesting for a route
\end{itemize}

\begin{figure}[htbp]
	\centering
	\includegraphics[scale = 0.5]{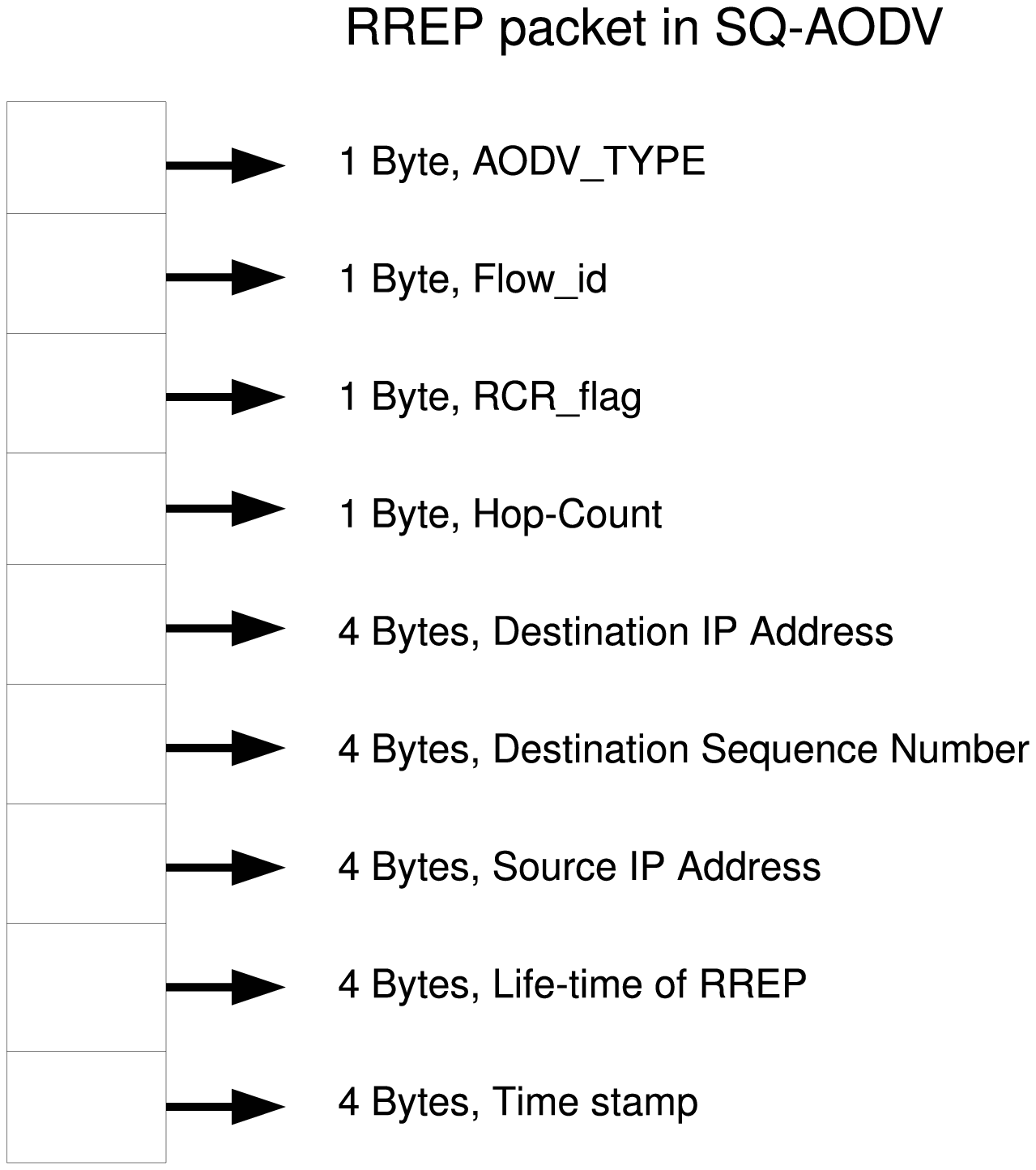}
	\caption{RREP Packet Format in SQ-AODV }
	\label{fig424}
\end{figure}

The RREP packet format modified for SQ-AODV is as shown in Figure~\ref{fig424} and the fields added are: 
\begin{itemize}
\item \textbf{Flow$\_$id} -- A sequence number uniquely identifying a flow (for which a route is desired) when taken in conjunction with the originator and the destinations IP address

\item \textbf{RCR$\_$flag} -- A flag to indicates whether the RREP packet is a Route Change Request (RCR) packet or not
\end{itemize}

Flow$\_$id field is added to facilitate the per-session based routing in SQ-AODV. The make-before-break re-routing mechanism of SQ-AODV requires an RCR packet to be sent by a node which is about to drain. RREP packet is used as RCR packet by setting RCR$\_$flag to $1$. The IP address of the node about to drain is sent in place of destination sequence number field of RREQ, this information is critical for handling the RCR packet in the network.

\begin{figure}[htbp]
	\centering
	\includegraphics[scale = 0.5]{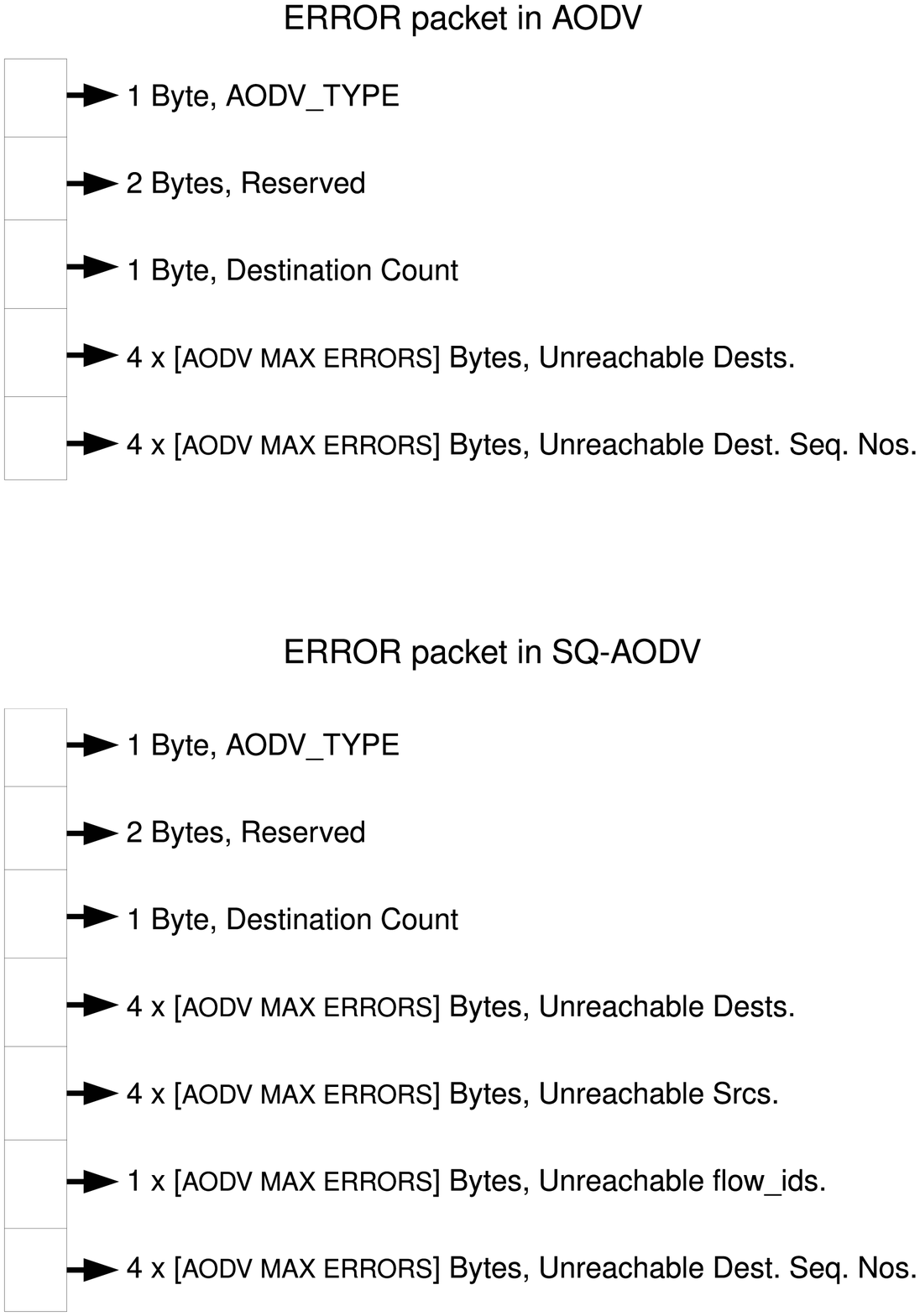}
	\caption{ERROR Packet Format in AODV and SQ-AODV}
	\label{fig425}
\end{figure}

The format of the ERROR packet both in AODV and SQ-AODV is as shown in the Figure~\ref{fig425}, and contains the following fields respectively:
\begin{itemize}
\item \textbf{Type} -- Indicates the type of the packet to differentiate between RREQ, RREP, ERROR packets

\item \textbf{Destcount} -- The number of unreachable destinations included in the ERROR packet

\item \textbf{Unreachable destinations} -- The IP address of the destinations that have become unavailable due to a link break

\item \textbf{Unreachable destination sequence number} -- The sequence number in the route table entry for the destination listed in the Unreachable destination field

\end{itemize}

The following fields are added to the ERROR packet format of AODV to use it for SQ-AODV:
\begin{itemize}
\item \textbf{Unreachable source} -- The IP address of the source, which is part of the triple (src/dst/flow$\_$id) identifying a route, got affected due to link failure

\item \textbf{Unreachable flow$\_$id} -- The flow$\_$id of a flow, which is part of the triple (src/dst/flo-\\w$\_$id) identifying a route, got affected due to link failure
\end{itemize}

Since the routing in SQ-AODV is per session based,  source and flow$\_$id information has to be carried as additional information by the ERROR packets to facilitate the route maintenance, when a link break due to either mobility of a node or variation in the medium occurs.

\newpage

\section{Route Discovery in SQ-AODV}

\begin{figure}[htbp]
	\centering
	\includegraphics[scale = 0.5]{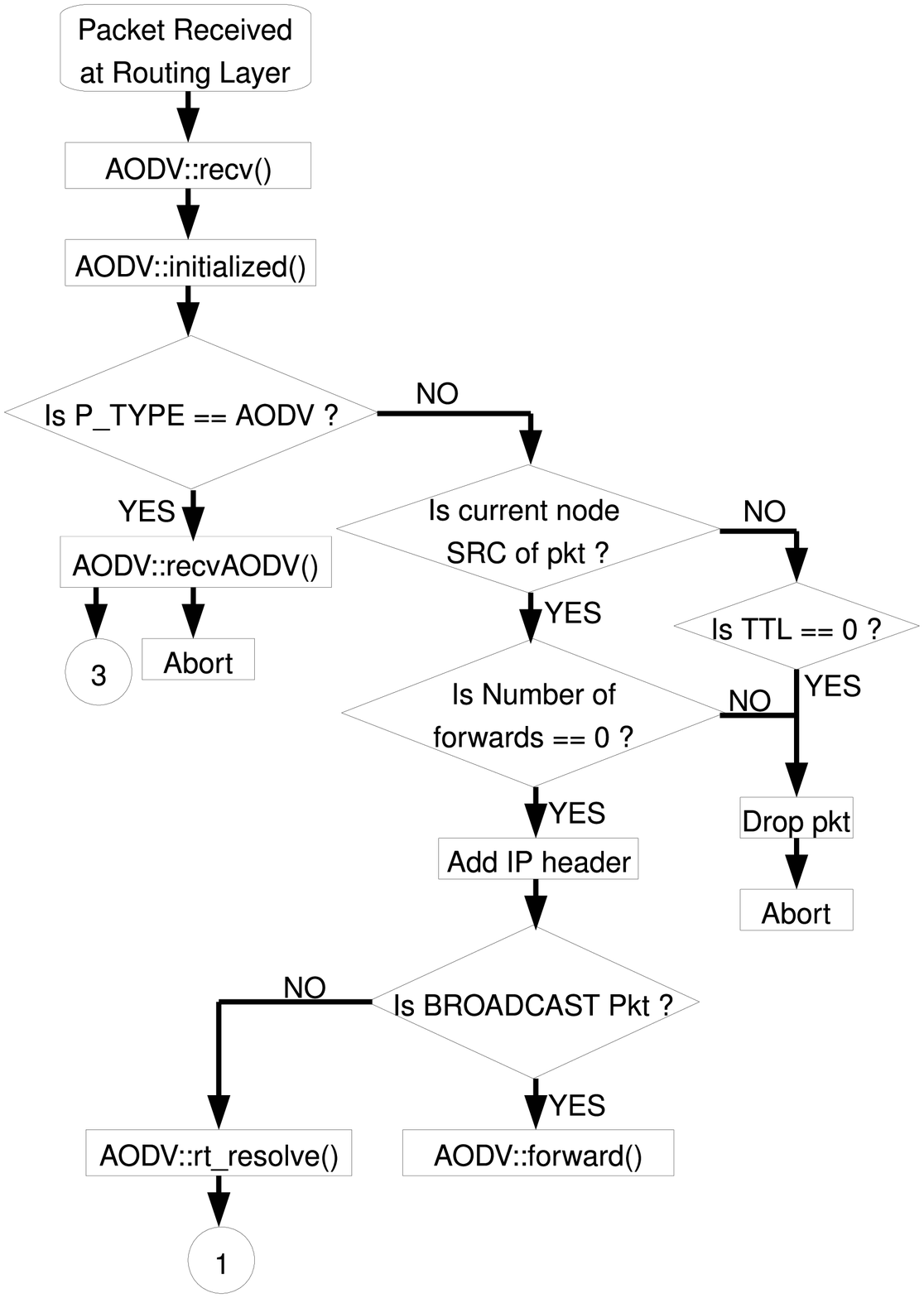}
	\caption{Flow-chart of AODV::recv() Function}
	\label{fig43}
\end{figure}

When a source node wants to communicate with a destination, the application layer of the source node generates data packets and send them to the network layer. In AODV or SQ-AODV a packet is received by AODV::recv(Packet *p, Handler *h) function. The basic operation of the recv function is depicted in Fig.~\ref{fig43}.

\begin{figure}[htbp]
	\centering
	\includegraphics[scale = 0.5]{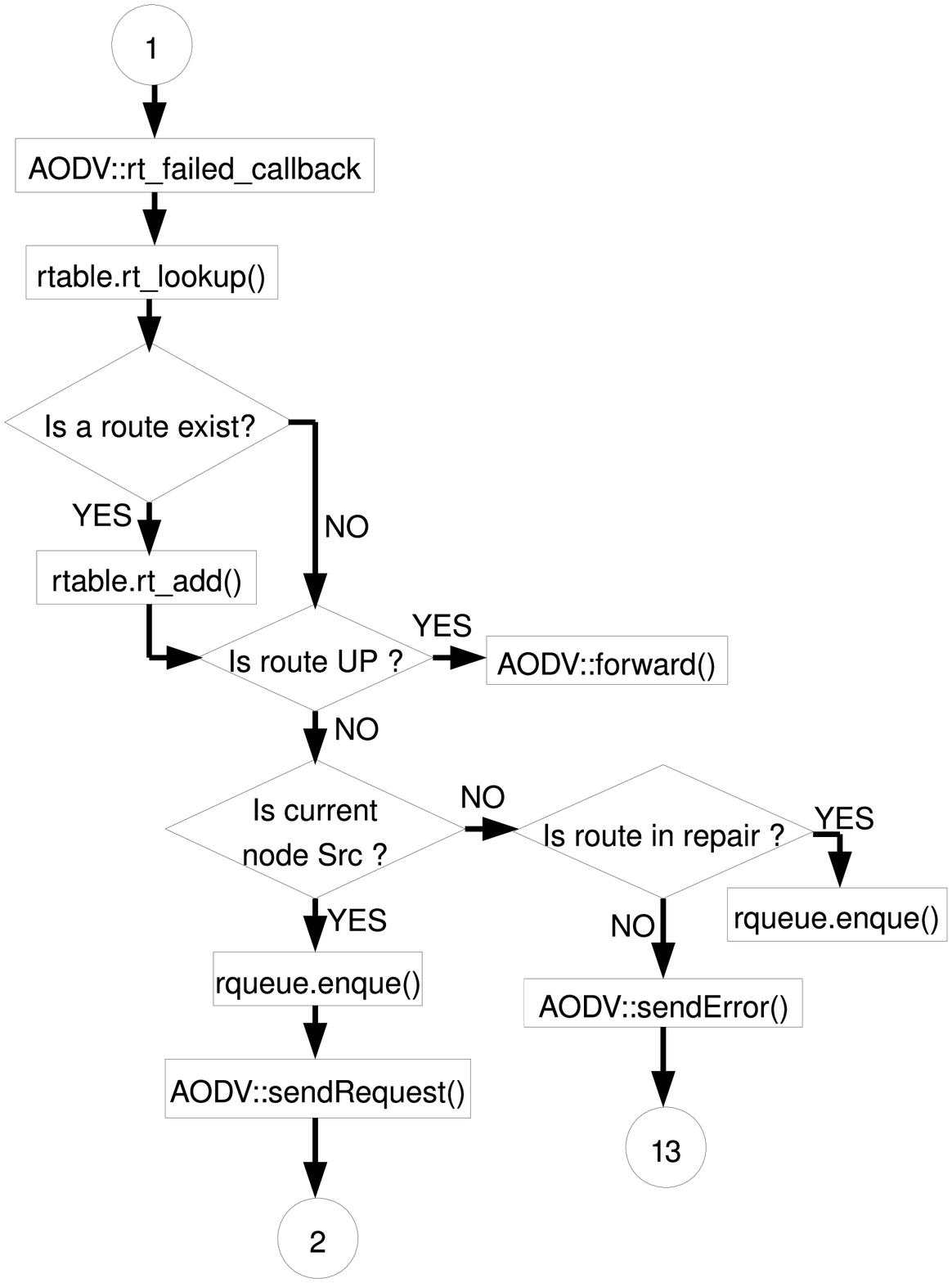}
	\caption{Flow-chart of AODV::rt$\_$resolve() Function}
	\label{fig44}
\end{figure}

The source node checks whether a valid route to the destination exists (in SQ-AODV, a route is identified by triple src/dst/flow$\_$id). If a valid route at the source exists then data packets are forwarded using this valid route. If a route does not exist the source node initiates a route discovery process. The function AODV::rt$\_$resolve(Packet *p) is responsible for resolving a route for the data packets and its operation is as depicted in Fig.~\ref{fig44}.

\begin{figure}[htbp]
	\centering
	\includegraphics[scale = 0.5]{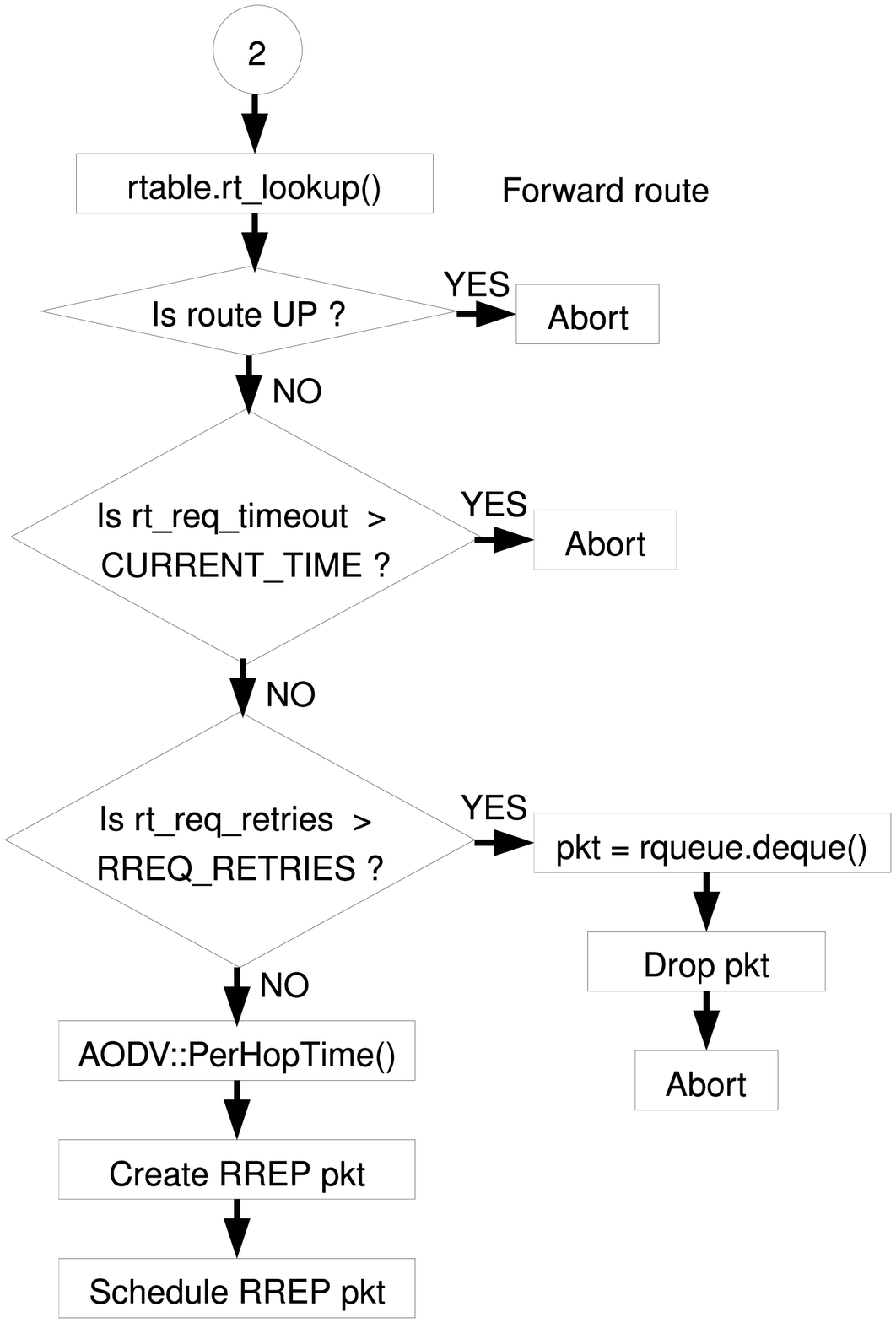}
	\caption{Flow-chart of AODV::sendRequest() Function}
	\label{fig45}
\end{figure}

A source node initiates a route discovery process by broadcasting a RREQ packet. As shown in Figure~\ref{fig422}, in SQ-AODV flow$\_$id and session-duration information (if specified) is carried in RREQ packet to facilitate a per session based routing, and find routes that last for the duration of a session. AODV::sendRequest(nsaddr$\_$t dst) function (AODV::sendR-\\equest(nsaddr$\_$t src, nsaddr$\_$t dst, u$\_$int8$\_$t flow$\_$id) function in SQ-AODV) is responsible for broadcasting RREQ packets to 1-hop neighbors, and its operation is depicted in Figure~\ref{fig45}.

\begin{figure}[htbp]
	\centering
	\includegraphics[scale = 0.5]{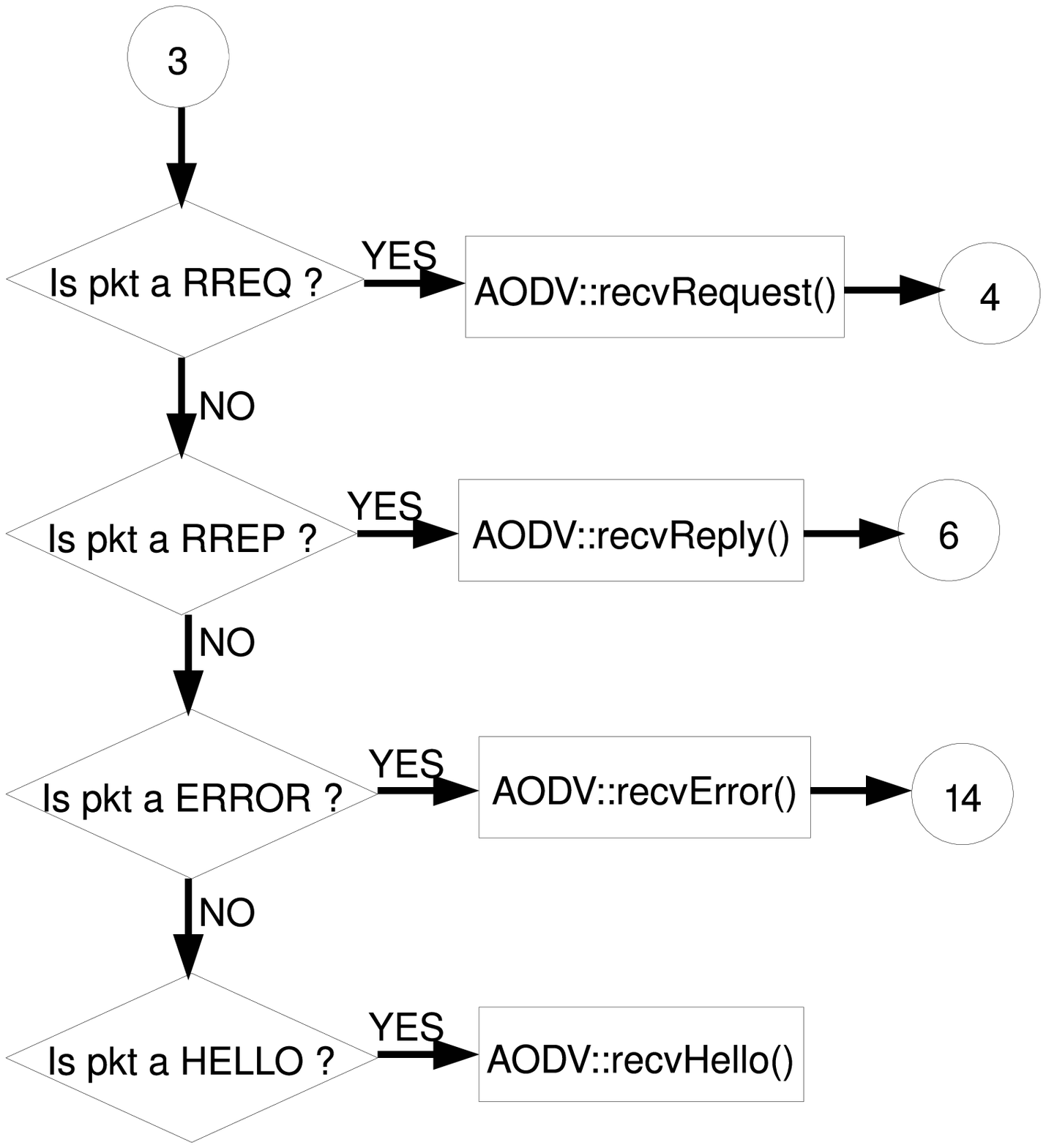}
	\caption{Flow-chart of AODV::recvAODV() Function}
	\label{fig46}
\end{figure}

\begin{figure}[htbp]
	\centering
	\includegraphics[scale = 0.5]{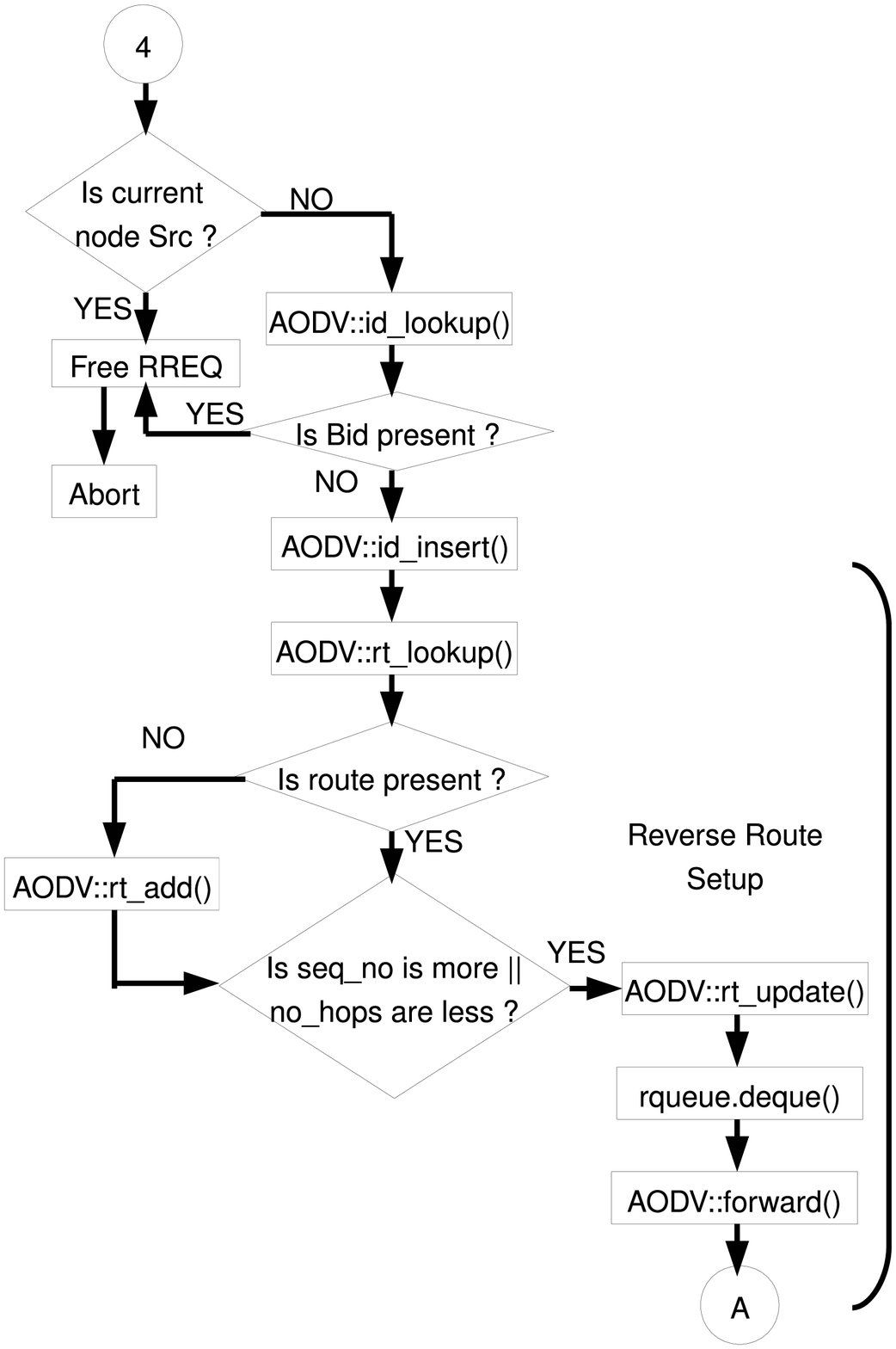}
	\caption{Flow-chart of AODV::recvRequest() Function}
	\label{fig47}
\end{figure}

\begin{figure}[htbp]
	\centering
	\includegraphics[scale = 0.5]{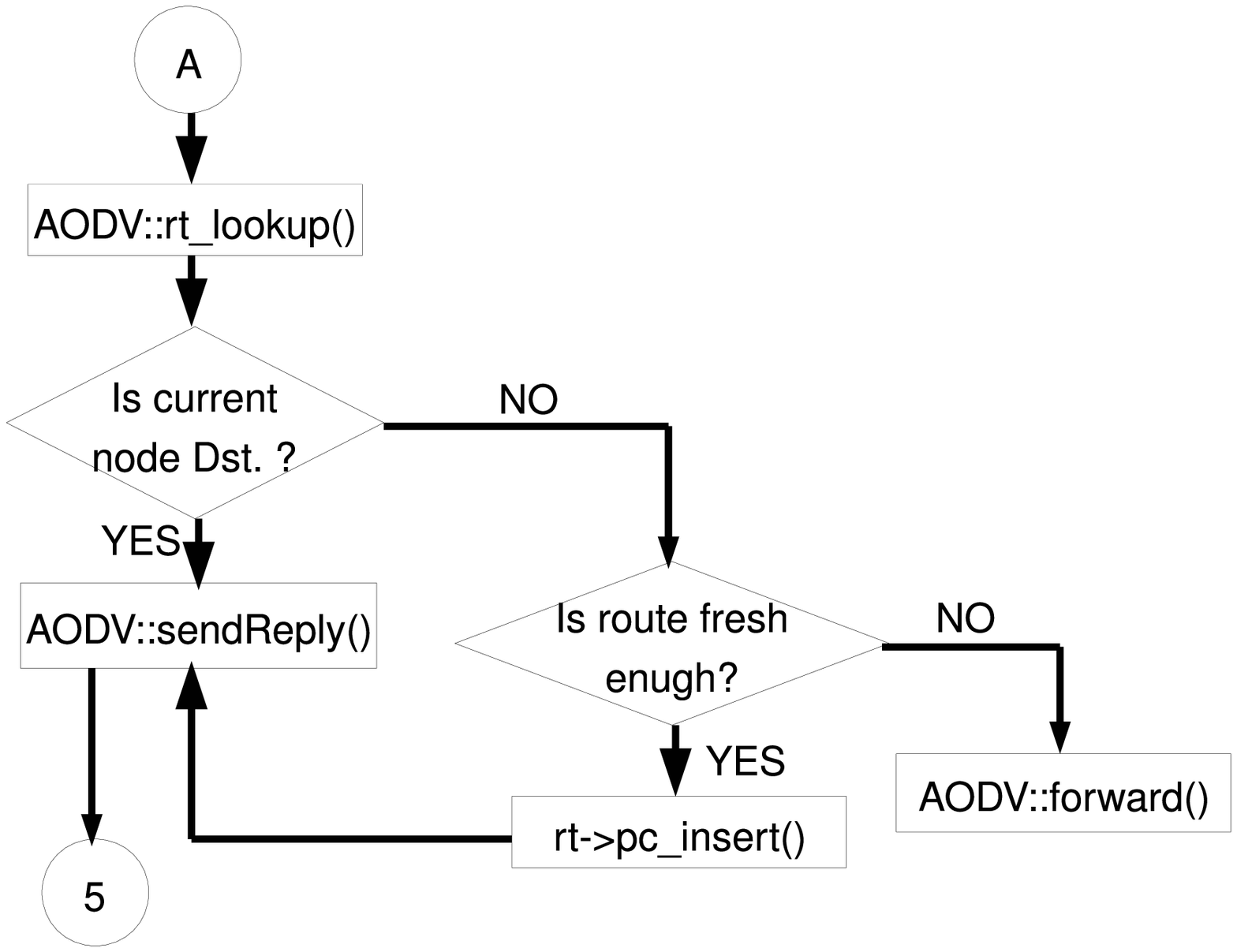}
	\caption{Flow-chart of AODV::recvRequest() Function}
	\label{fig47_A}
\end{figure}

When an intermediate node receives a RREQ packet the packet is processed by the AODV::recvAODV(Packet *p) function and AODV::recvRequest(Packet *p) functions, whose operation is as depicted in Figure~\ref{fig46}, and Figures~\ref{fig47} and~\ref{fig47_A} respectively. In case of SQ-AODV an intermediate node cannot reply to the RREQ packets because, (i) the routing is per session based (ii) destination selects a path (among available) which maximizes the life-time of the route (iii) applies the Algorithm 1 explained in Chapter~\ref{intro-sq-aodv} to forward the RREQ packets. Each intermediate node both in AODV and SQ-AODV sets up a reverse route to the previous hop from which the RREQ packet has been received. In SQ-AODV the bottleneck life-time field of RREQ packet is updated before re-broadcasting the RREQ packet.

\begin{figure}[htbp]
	\centering
	\includegraphics[scale = 0.5]{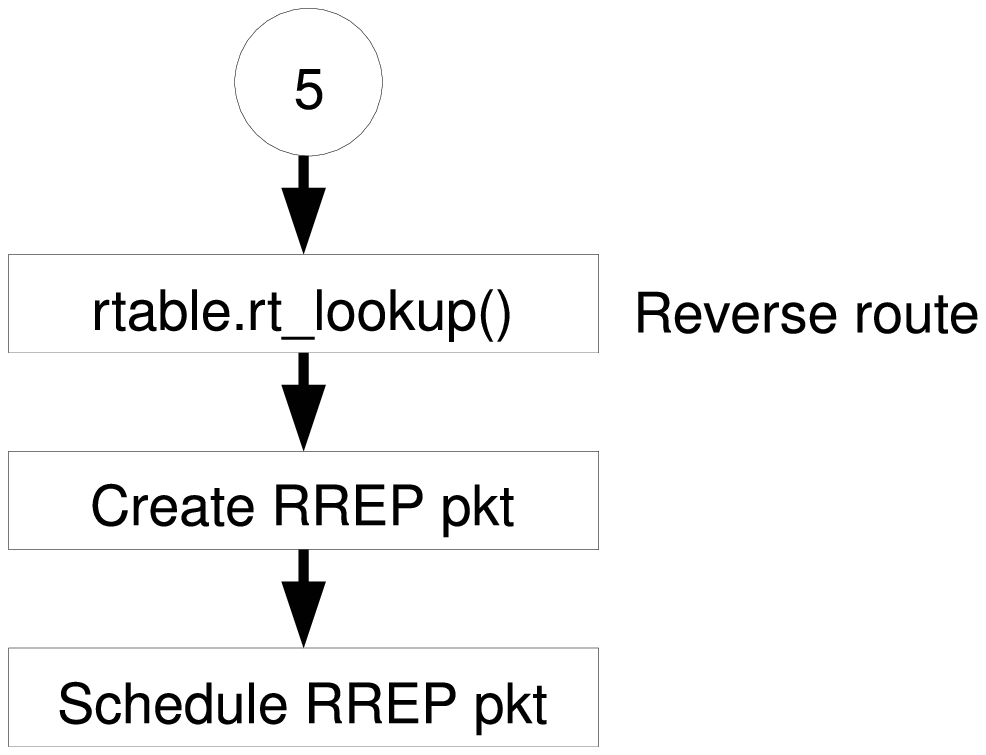}
	\caption{Flow-chart of AODV::sendReply() Function}
	\label{fig48}
\end{figure}

\begin{figure}[htbp]
	\centering
	\includegraphics[scale = 0.5]{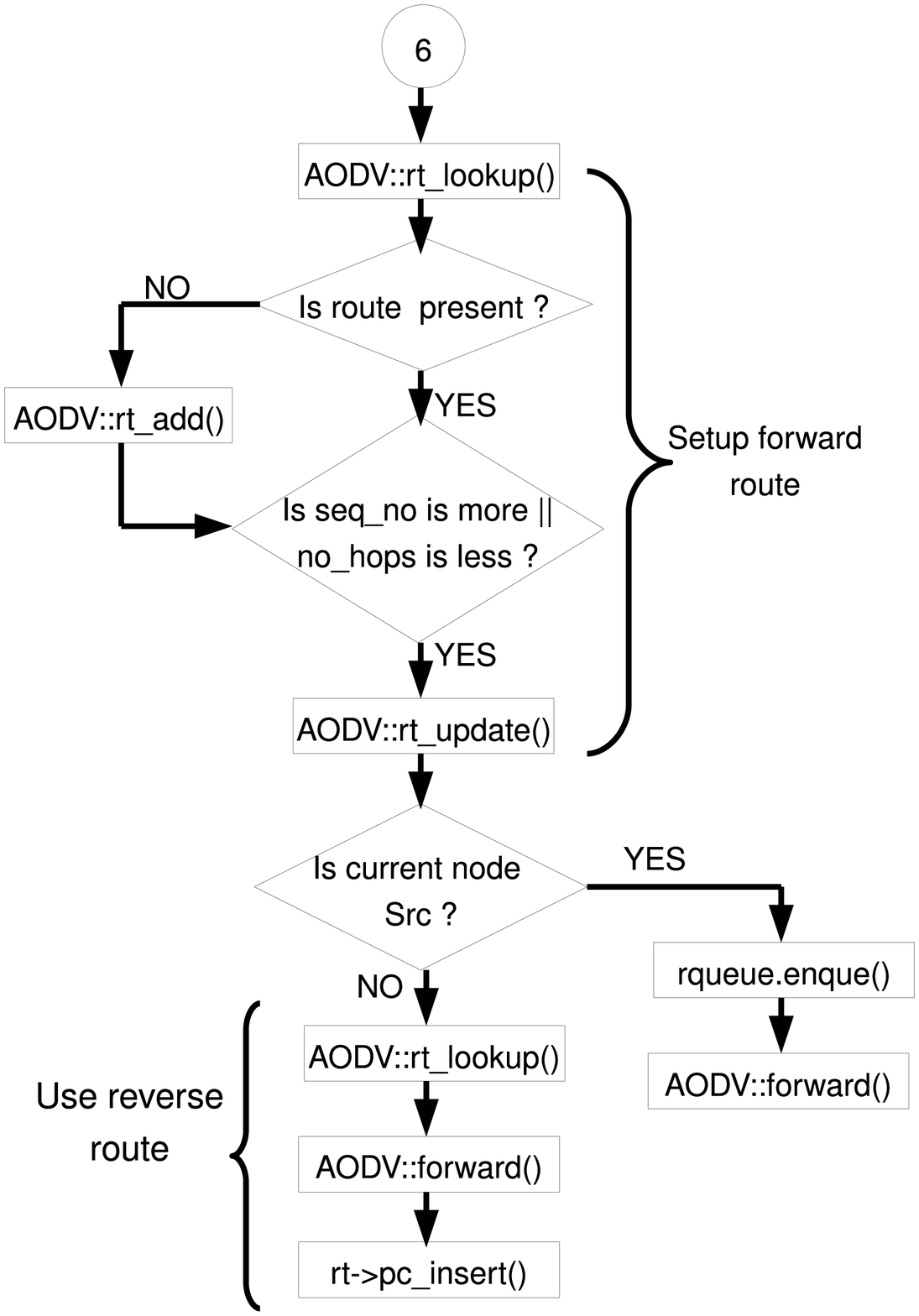}
	\caption{Flow-chart of AODV::recvReply() Function}
	\label{fig49}
\end{figure}

In AODV, when a RREQ packet reaches the destination, node generates a RREP packet to reply to this request. On the other hand in SQ-AODV, a destination node waits for either 0.25 seconds after the first RREQ arrival or three RREQ packet arrivals from different paths to reply to a RREQ that maximizes life-time of the path. Figure~\ref{fig48} depicts the basic operation of AODV::sendReply(nsaddr$\_$t ipdst, u$\_$int32$\_$t hop$\_$count, nsaddr$\_$t rpdst, u$\_$int32$\_$t rpseq, u$\_$int32$\_$t lifetime, double  timestamp) function (AODV::sendReply(nsaddr$\_$t ipdst, u$\_$int32$\_$t hop$\_$count, nsaddr$\_$t rpdst, u$\_$int32$\_$t rpseq, u$\_$int32$\_$t lifetime, double timestamp, u$\_$int8$\_$t flow$\_$id, u$\_$int8$\_$t rcr$\_$flag) function in SQ-AODV) in AODV. An intermediate node receiving this RREP in AODV or SQ-AODV will unicast the RREP packet towards the source using reverse route information, and sets a forward pointer to the hop from which the RREP has arrived. This forward pointer is nothing but the route established for the current session. The RREP packet received by a node is processed using the AODV::recvReply(Packet *p), and its operation is depicted in Figure~\ref{fig49}. The source start forwarding data packets once it receives the RREP packet.

\newpage

\section{Route Maintenance in SQ-AODV}
AODV has capabilities of both proactive and reactive route maintenance. Since SQ-AODV is an enhancement to AODV it inherits the capabilities of AODV. Apart from these route maintenance capabilities, SQ-AODV has a mechanism of make-before-break re-routing. We first explain the basic AODV route maintenance mechanisms and then give the make-before-break re-routing feature of SQ-AODV.

The proactive maintenance in AODV is by periodic hello messages in which every node in the network broadcast hello messages to its neighbors every one second. Reactive maintenance is done by sending ERROR packets to the sources whenever the link layer detects a link breakage. If the reactive maintenance is enabled then proactive maintenance (NeighborTimer and HelloTimer) is disabled. 

\begin{figure}[htbp]
	\centering
	\includegraphics[scale = 0.5]{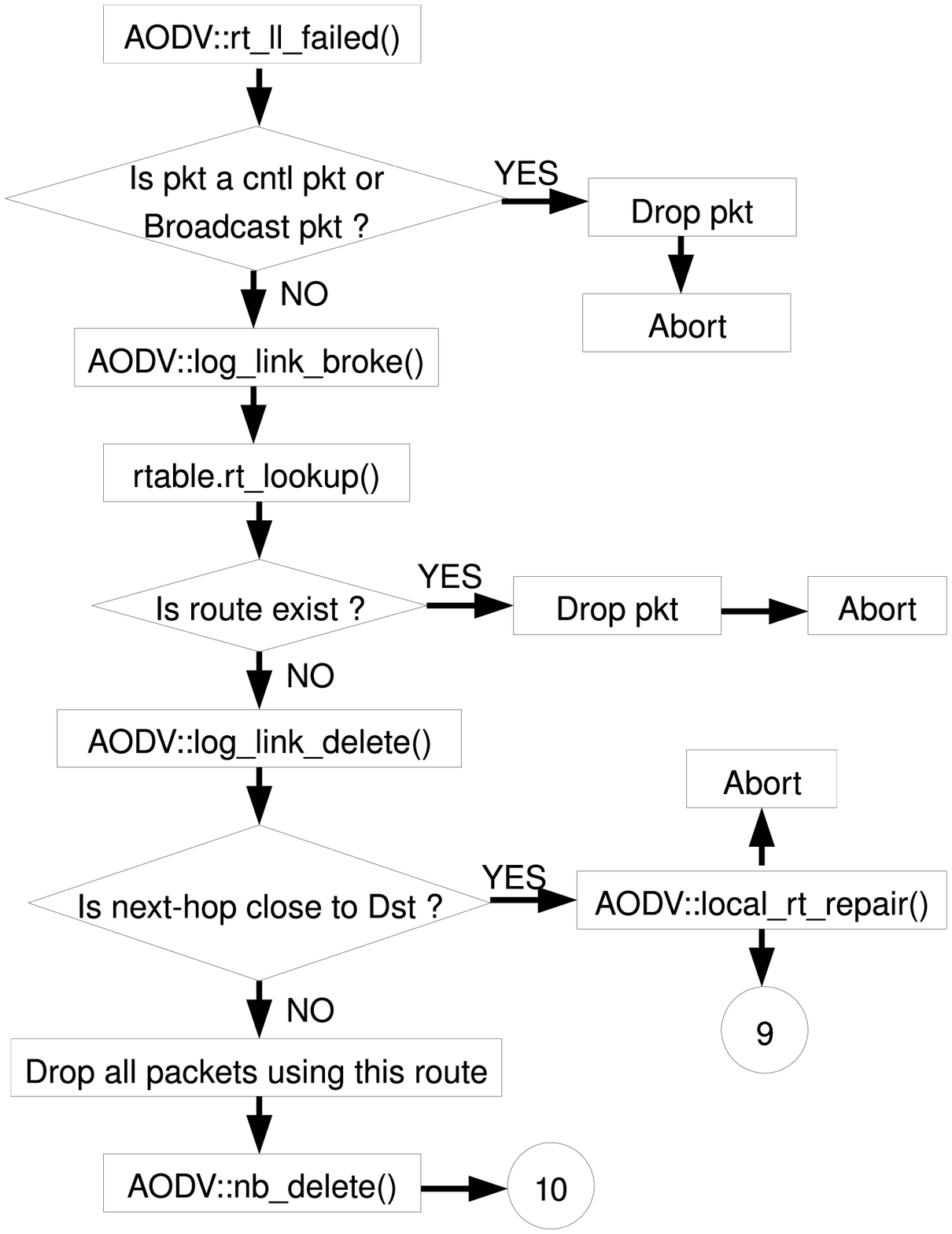}
	\caption{Flow-chart of AODV::rt$\_$ll$\_$failed() Function}
	\label{fig410}
\end{figure}

\begin{figure}[htbp]
	\centering
	\includegraphics[scale = 0.5]{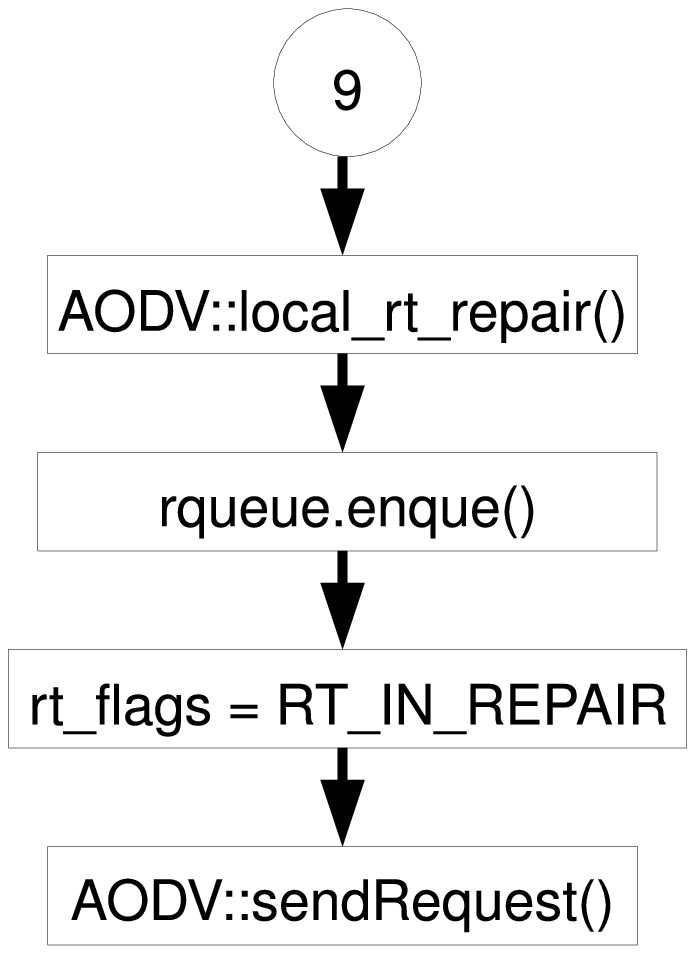}
	\caption{Flow-chart of AODV::local$\_$rt$\_$repair() Function}
	\label{fig413}
\end{figure}

When a node detects a link failure through link layer feedback it initiates the maintenance process by broadcasting the ERROR packet to 1-hop neighbors.
The AODV::rt$\_$ll$\_$f-ailed(Packet *p) function is responsible for the initiation of the maintenance process and its operation as depicted in Figure~\ref{fig410}. During a route maintenance the route is repaired either through a local repair (if the broken link is near to the destination) or through the ERROR packets. Local route repair is done using the AODV::local$\_$rt$\_$repair(aodv$\_$rt$\_$entry *rt, Packet *p) function, and its basic operation is as depicted in Figure~\ref{fig413}

In case of the broken link close to the source node the route maintenance is initiated by the AODV::handle$\_$link$\_$failure(nsaddr$\_$t id) function whose operation is depicted in Figure~\ref{fig415}. The handle$\_$link$\_$failure function sends an ERROR packet with the help of AODV::sendError(Packet *p, bool jitter=true) function and the basic operation of sendError function is depicted in Figure~\ref{fig417}. when the ERROR packet finally reaches the source of a flow whose route is down due to the link failure, the source node initiates a route discovery process to find an alternate route.

\subsection{Make-Before-Break Re-routing in SQ-AODV}
In this section we explain the details of make-before-break re-routing feature of SQ-AODV. To implement make-before-break re-routing mechanism, we have added a timer called RCR-timer in the basic AODV code. The RCR-timer expires every 100 msec, and on every expiry the AODV::rcr() function is called. The combined operation of RCR-timer and rcr function is as depicted in Fig.~\ref{fig419}. In rcr function a check is performed to find whether the current-energy of the node is less than \textbf{Threshold-2}. If so, a Route Change Request (RCR) packet is broadcast to the 1-hop neighbors of this node.

After receiving a broadcast RCR packet, each neighbor checks whether the drained node is the destination or the source or an intermediate node of a route. This check is performed for every forward entry in the routing table, and depending on whether drained node is source/destination/intermediate node for the flows of neighbors, each neighbor processes the RCR packet uniquely as depicted in Fig.~\ref{fig420}.

If the drained node is, a destination of any flow in the routing table of the neighbor indicating its destination has drained, and the neighboring node which received the broadcast RCR is a source of the flow then the source node will immediately stop the traffic. On the other hand, if the neighboring node which received the broadcast RCR is an intermediate node of the flow then this intermediate node sets the RCR-flag in the routing table and sends an unicast RCR packet towards the source of the flow.

If the node from which a broadcast RCR has come happens to be a source for any of the routes in the routing table of neighbors then the neighboring node simply free this broadcast RCR packet.

If the node from which a broadcast RCR has come happens to be an intermediate node for any of the routes in the routing table of neighbors, and if the 1-hop neighboring node is a source of the flow then the source broadcast a RREQ packet to initiate a route discovery process. On the other hand, if the node which received the broadcast RCR is an intermediate node of a flow then this intermediate node sets the RCR-flag in the routing table and sends an unicast RCR packet towards the source of the flow.

In case of a route with more than one hop an intermediate node has to unicast the RCR packet towards the source of the flow. Now, if this unicasted RCR packet has reached one more intermediate node of the flow then the intermediate node forwards the RCR packet using the reverse route information in its routing table. On the other hand if the unicasted RCR has reached a source node of a flow the source node will check whether the drained node is the destination or an intermediate node of the flow, if the destination has drained then, the source simply stops the traffic, else if an intermediate node has drained then source initiates a route discovery process by broadcasting a RREQ packet.

\begin{figure}[htbp]
	\centering
	\includegraphics[scale = 0.5]{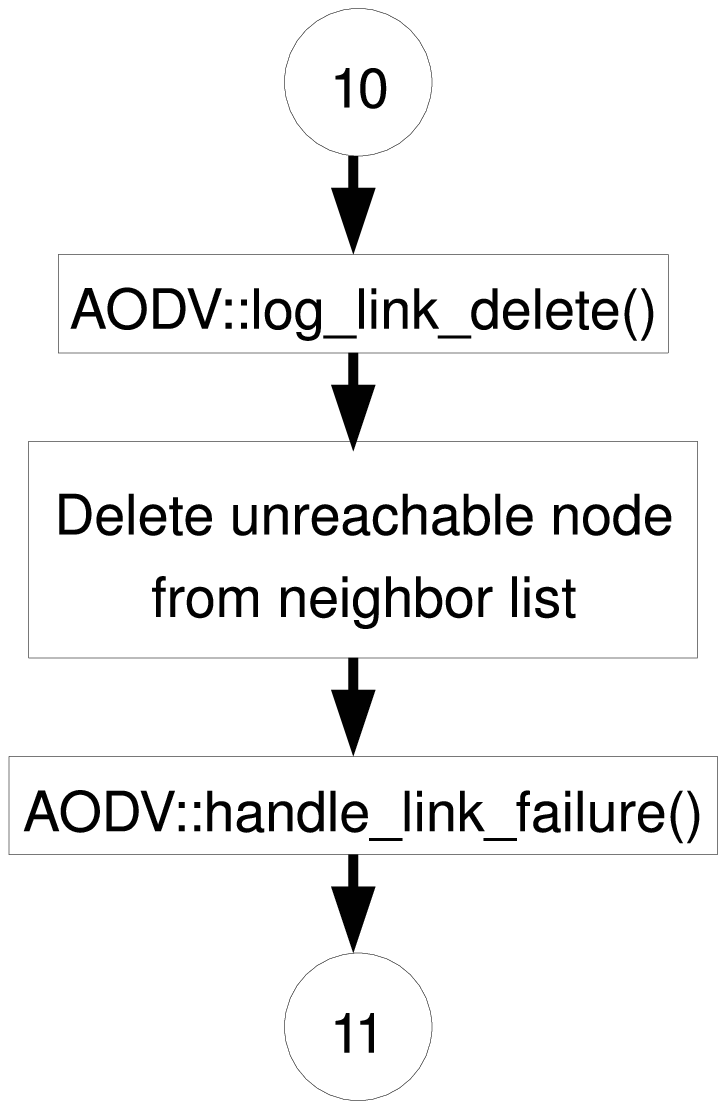}
	\caption{Flow-chart of AODV::nb$\_$delete() Function}
	\label{fig414}
\end{figure}

\begin{figure}[htbp]
	\centering
	\includegraphics[scale = 0.5]{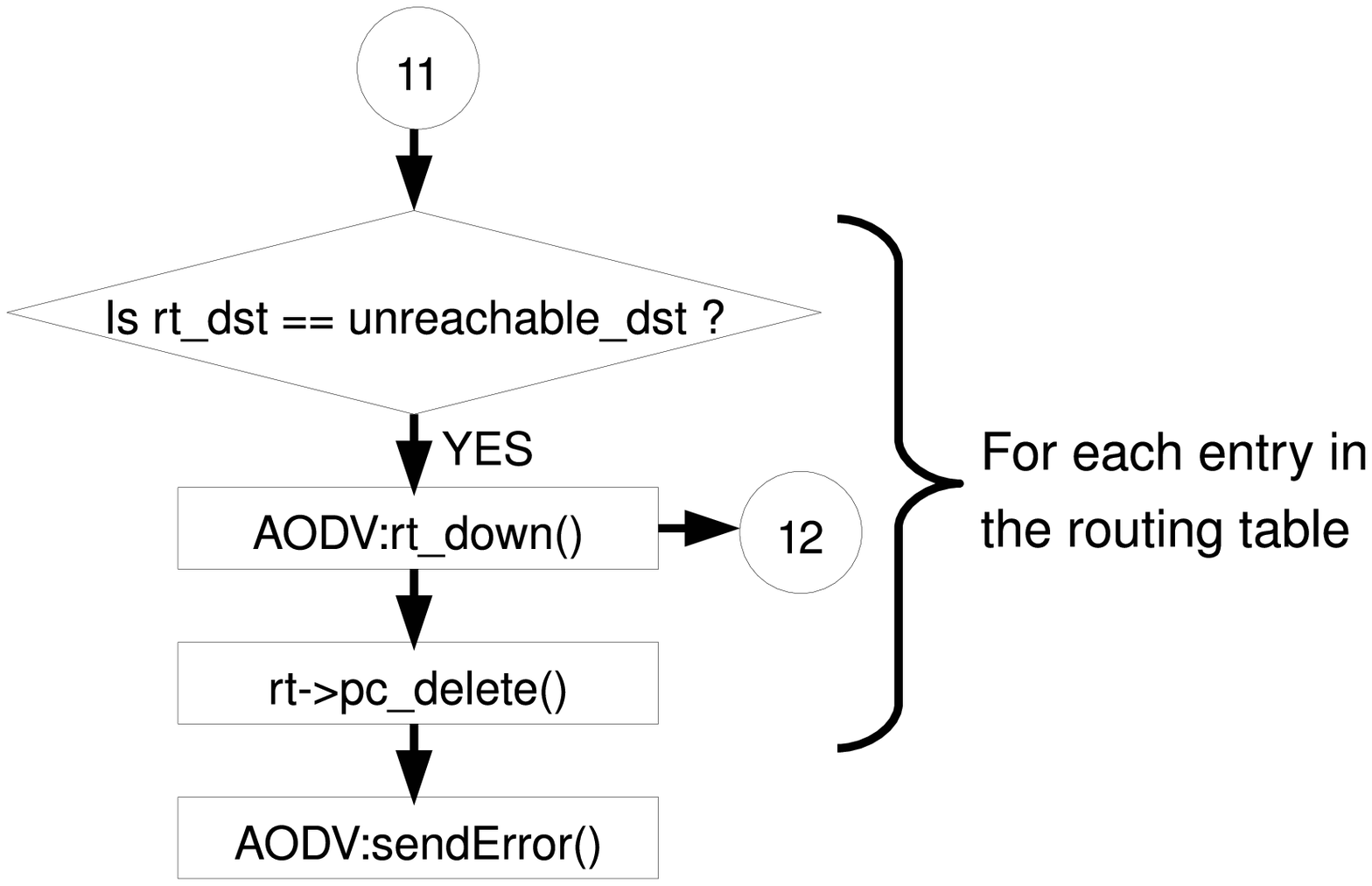}
	\caption{Flow-chart of AODV::handle$\_$link$\_$failure() Function}
	\label{fig415}
\end{figure}

\begin{figure}[htbp]
	\centering
	\includegraphics[scale = 0.5]{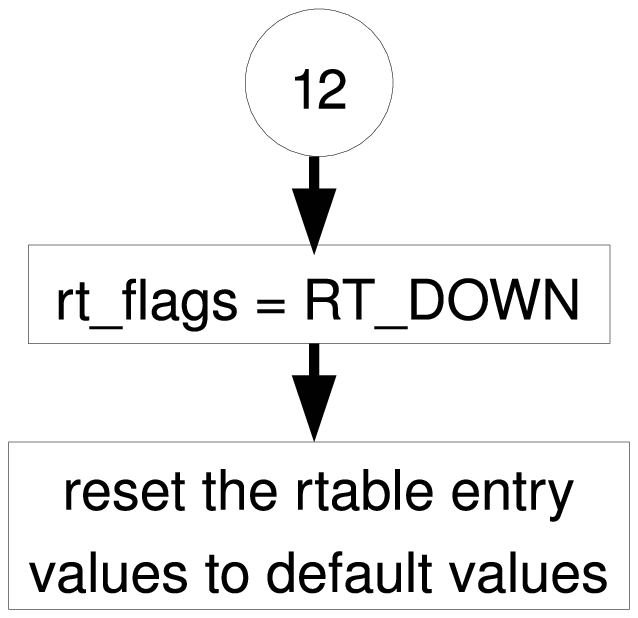}
	\caption{Flow-chart of AODV::rt$\_$down() Function}
	\label{fig416}
\end{figure}

\begin{figure}[htbp]
	\centering
	\includegraphics[scale = 0.5]{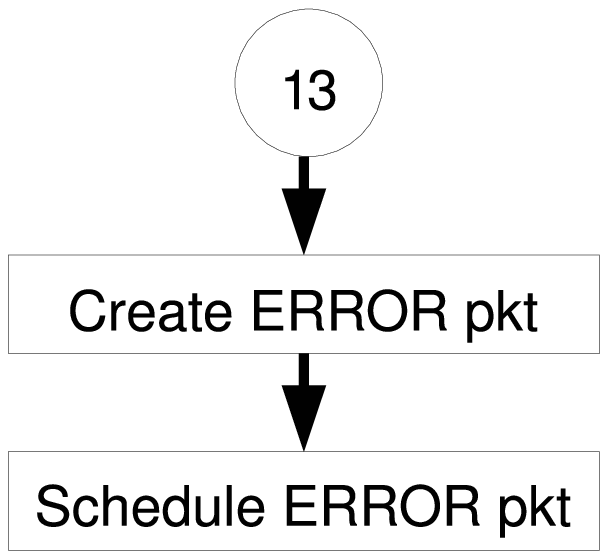}
	\caption{Flow-chart of AODV::sendError() Function}
	\label{fig417}
\end{figure}

\begin{figure}[htbp]
	\centering
	\includegraphics[scale = 0.5]{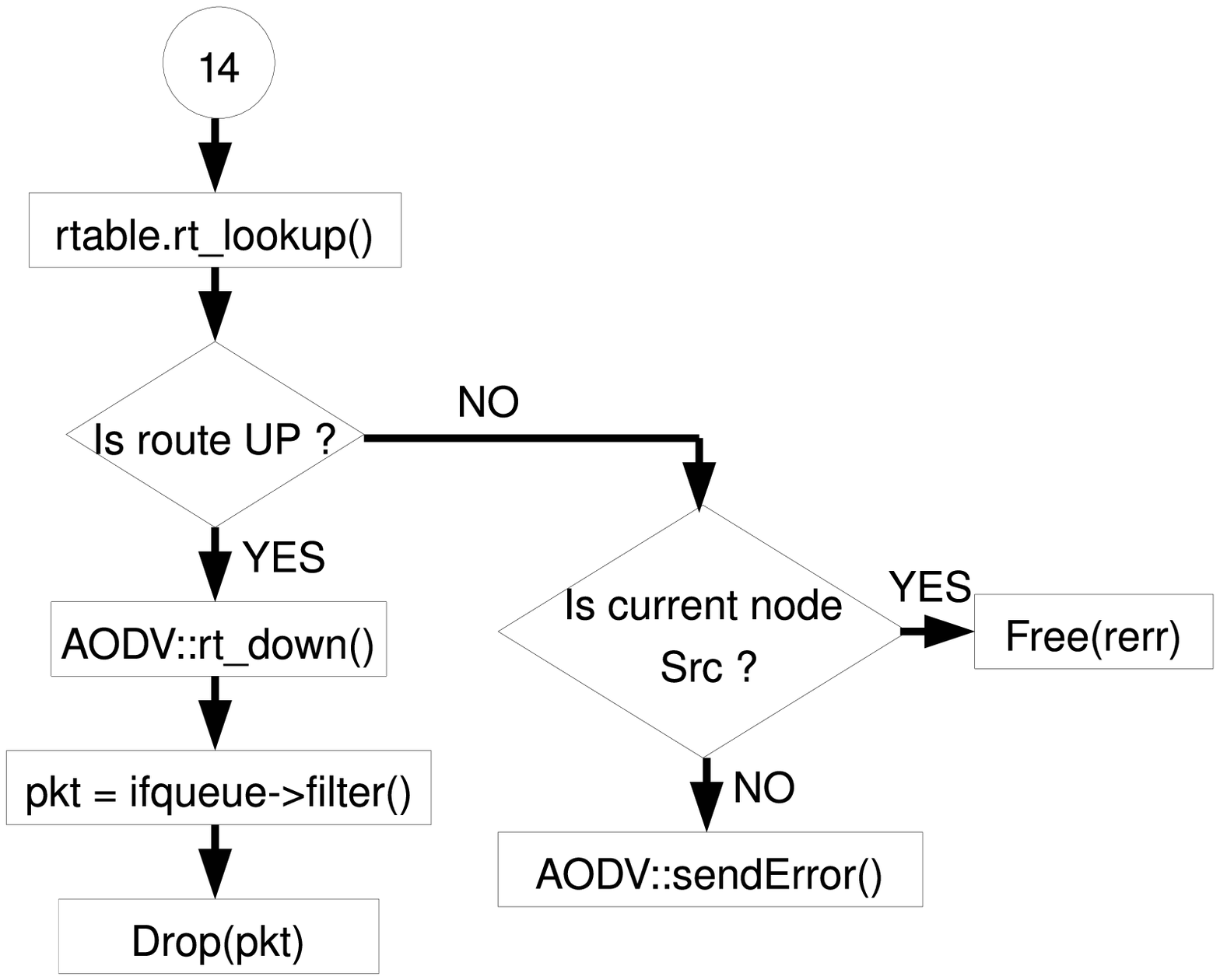}
	\caption{Flow-chart of AODV::recvError() Function}
	\label{fig418}
\end{figure}

\begin{figure}[htbp]
	\centering
	\includegraphics[scale = 1.0]{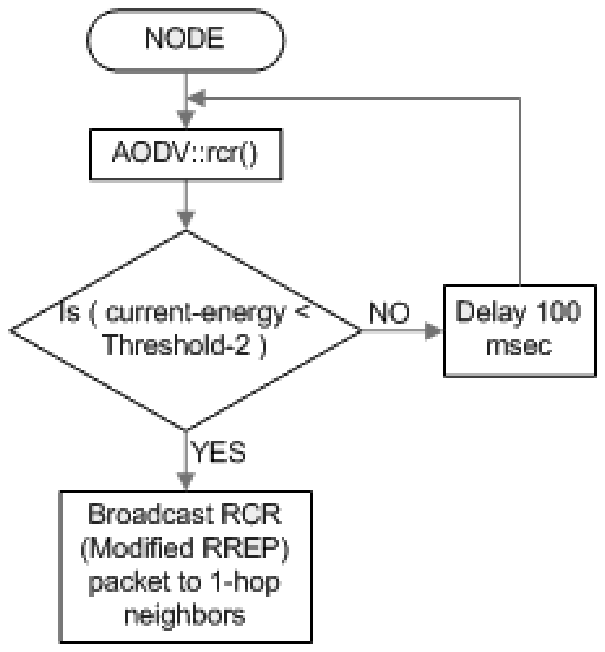}
	\caption{Combined Flow-chart of RCR-timer and RCR function of SQ-AODV}
	\label{fig419}
\end{figure}

\begin{figure}[htbp]
	\centering
	\includegraphics[scale = 1.0]{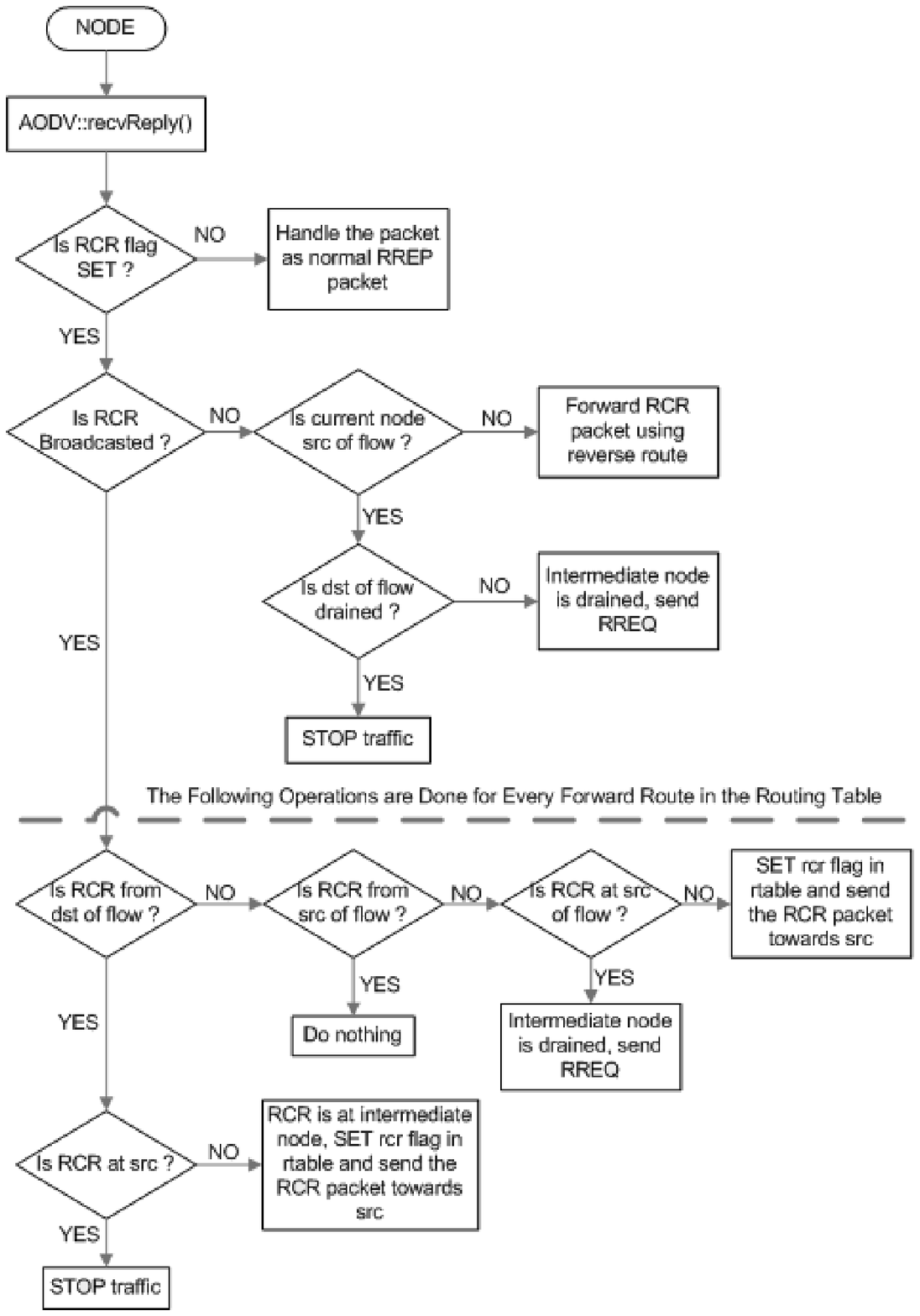}
	\caption{Flow-chart of Make-before-break Feature of SQ-AODV}
	\label{fig420}
\end{figure}